%% file: main.tex
\documentclass{article}
\usepackage{amsmath}
\usepackage{amssymb}
\usepackage{graphicx} 
\usepackage{subcaption}
\usepackage{epigraph} 
\usepackage{booktabs}
\usepackage{multirow}
\usepackage{comment}
\usepackage{soul}
\usepackage{url}
\usepackage{authblk}
\usepackage[normalem]{ulem}
\usepackage[table]{xcolor}
\usepackage[framemethod=TikZ]{mdframed}

\newtheorem{theorem}{Definition}[section]

\newcommand{\method}{\texttt{Apollonion}}
\newcommand{\Response}{\texttt{Response}}
\newcommand{\Reflect}{\texttt{Reflect}}

\newcommand{\Init}{\texttt{Profile Initialization}}

\newcommand{\Retrieve}{\texttt{Retrieve}}

\newcommand{\query}{\mathbf{Q}}
\newcommand{\response}{\mathbf{R}}
\newcommand{\profile}{\mathbf{P}}
\newcommand{\reflection}{\mathbf{Ref}}
\newcommand{\selectedProfile}{\mathbf{\hat{P}}}
\newcommand{\conversation}{\mathbf{H}}

\newcommand{\diag}{\mathbf{M}_{\theta}}

\newcommand{\PersonalizedMetric}{\text{Pers}}

\begin{document}

\title{\method{}: Profile-centric Dialog Agent}
\author[1]{Shangyu Chen}
\author[1]{Zibo Zhao}
\author[1]{Yuanyuan Zhao}
\author[1]{Xiang Li}
\affil[1]{Meituan, Beijing, China}
\affil[ ]{\{chenshangyu03,zhaozibo02,zhaoyuanyuan09,lixiang245\}@meituan.com}
\date{}

\maketitle
\epigraph{Know yourself}{}

\begin{figure}[!htbp]
    \centering
    \includegraphics[width=0.5\textwidth]{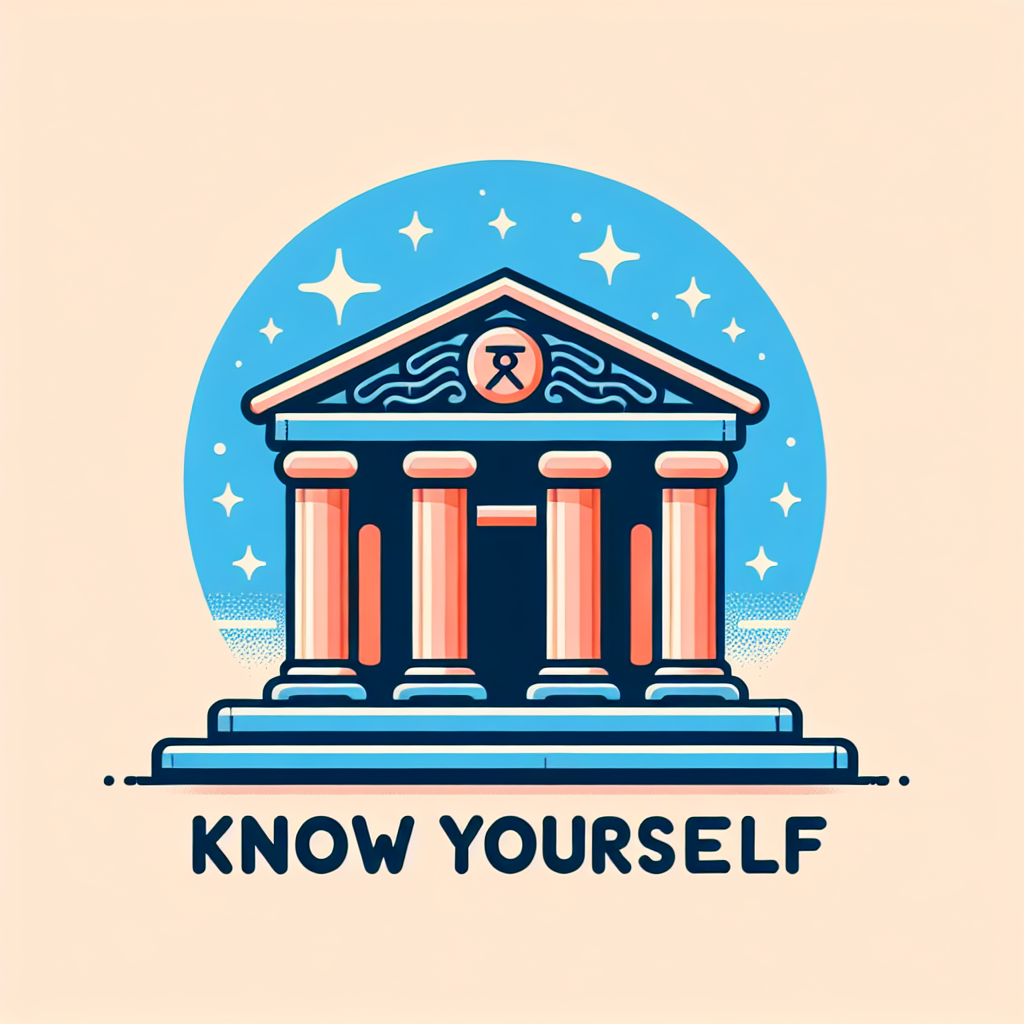}
    \caption{``Know yourself'' is a philosophical maxim which was inscribed upon the Temple of Apollo (also known as \method{}) in the ancient Greek precinct of Delphi. 
    It becomes famous and survive today for it is believed that the great philosopher Socrate cite it as creed: The most important thing in life is to gain understanding of oneself.}
    \label{fig:logo}
\end{figure}

\section{Abstract}
The emergence of Large Language Models (LLMs) has innovated the development of dialog agents. Specially, a well-trained LLM, as a central process unit, is capable of providing fluent and reasonable response for user's request. 
Besides, auxiliary tools such as external knowledge retrieval, personalized character for vivid response, short/long-term memory for ultra long context management are developed, completing the usage experience for LLM-based dialog agents. 
However, the above-mentioned techniques does not solve the issue of \textbf{personalization from user perspective}: agents response in a same fashion to different users, without consideration of their features, such as habits, interests and past experience. In another words, current implementation of dialog agents fail in ``knowing the user''. The capacity of well-description and representation of user is under development.
In this work, we proposed a framework for dialog agent to incorporate user profiling (initialization, update): user's query and response is analyzed and organized into a structural user profile, which is latter served to provide personal and more precise response. 
Besides, we proposed a series of evaluation protocols for personalization: to what extend the response is personal to the different users. 

The framework is named as \method{}, inspired by inscription of ``Know Yourself'' in the temple of Apollo (also known as \method{}) in Ancient Greek.
Few works have been conducted on incorporating personalization into LLM, \method{} is a pioneer work on guiding LLM's response to meet individuation via the application of dialog agents, with a set of evaluation methods for measurement in personalization.

\section{Introduction}
\input{content/introduction}

\section{Related Works}\label{sec:related-works}
\input{content/related-works}

\section{Method}\label{sec:method}
\input{content/method}

\section{Experiments}\label{sec:experiments}
\input{content/experiments}

\section{Conclusion}\label{sec:conclusion}
In this paper, we proposed a framework to deal with the challenge of user information incorporation for dialog agents to provide personal response. 
Specially, for each user, we construct an initial user profile (Sec.\ref{sec:method:init}) by collection of user interaction in Meituan and inferred by LLM. 
The the user profile and historical conversation are retrieved (Sec.\ref{sec:method:retrieve}) according to user query, followed by a response module (Sec.\ref{sec:method:response}) to provide personal responses.
Finally, a reflection module (Sec.\ref{sec:method:reflect}) is set up to infer user's preference, recent activities according to the query, forming a detailed profile for latter response.

Besides, we proposed a series of evaluation protocols to quantify ``Personalization'' (Sec.\ref{sec:experiments:personalized-response}) and measure the performance of modules. 
The protocols are cheap and scalable, instead of relying third-party LLM for evaluation.

\clearpage

\bibliographystyle{plain}
\bibliography{references}

\clearpage

\section{Appendix}\label{sec:app}
\input{content/appendix}

\end{document}

%% file: content/introduction.tex
Agents is now becoming a promising direction for next-generation Artificial Intelligence (AI) applications based on the unprecedented capacity of LLMs for human language understanding. 
In this paper, we define \emph{agents} as ``A system of connecting LLMs for solving a series of tasks''. Normally one LLM serves as central node for router and main computation, with some LLMs for auxiliary assistance". 
For example, dialog agents \cite{lee2023prompted},\cite{xu2022long} can be considered as a main LLM as response unit, with one LLM for user query understanding, one for external knowledge retrieval, one for relevant long-term memory matching. 
In task-solving, LLMs are widely used \cite{li2024camel}, \cite{yao2022react}, \cite{yang2023auto},\cite{shen2024hugginggpt}, \cite{wu2023autogen}: these works set up a framework to include several agents who are customizaible and conversable to each other to solve a specific task.
Agents is further considered as next-generation operating system \cite{ge2023llm}. Specially one LLM serves as Central Processing Unit (CPU) while others are responsible for file management, memory and etc.

In these attempts for utilization of agents, dialog agents is attracting more and more attention for the following reasons: 
1) It has a long history for research and deployment, which grants a foundation for its further utilization. 
2) Dialog agents can be easily integrated into current form of e-commercial applications (search, recommendation system \cite{zhang2024usimagent}, \cite{wu2023survey}, \cite{zhang2023generative}), leading to conversational search/recommendation system.
3) Personalized assistants is a promising way of human-AI interaction. Though it has been developed as key component in modern operating system, it previously focused on task solving while lacking ability for user-friendly interactions, which can be greatly enhanced by dialog agents.
4) Other forms of agents (planning \cite{valmeekam2024planning}, \cite{park2023generative}, task solving \cite{chen2023multi}) rely largely on agents' interactions, where conversations occupy a huge portion. Dialog process and understanding serves as a foundation techniques for agents.

The emergence of LLMs accelerates the improvement of dialog agents \cite{zhang2019dialogpt}: Broadly, empowering dialog agents with LLM can be separated into: 
1) Role-play: endowing agents with specific character for vivid interactions \cite{shanahan2023role}, \cite{lu2024large}.
2) Response with personalization: User information is deeply involved and reflected in agents' response and interaction, leading to personal response \cite{salemi2023lamp}, \cite{zhang2018personalizing}.
However, these implementation of dialog agents seldom take user's features into consideration. User information is solely represented in their queries, which are processed as a common component for prompt construction in response generation. Shot / long-term memory can be somehow considered as user's historical information, however, its usage aims at retrieving relevant memory for precision.
In summary, user understanding and corresponding profile, advantage of the information is less studied in previous works, leading to a situation that dialog agents know little about the user it is responding and assisting.
Though lacking of user feature does not affect reasonable response, it hinders to provide personalized and more precise answers to users. 
For example, on the awareness of user's diet preference to be non-spicy, food recommendation is more precise and user-friendly. With the information that user has kept hiking for the past several weekends, dialog agents should infer and generate the user profile that his or her interests includes outdoor activities, leading to items recommendation with similar topic such as camping and bicycling facilities.

Based on the observation, we proposed a framework for dialog agents that incorporate user profiling to provide response with individuation: 
A structural data is set as user profile, representing and storing the characters (including basic information, deduced interests and etc) of user. 
User profile is updated dynamically: an introduced LLM module is utilized to analyze and understand user's query, leading to reflections for the user. The LLM infers user' characters based on each of his or her query. 
For example, user's query for a specific topic (``Recommend a salad for lunch'') is regarded as a preference or interests (``This user prefers light food''). Moreover, the latter query may add more information (``What is good for sore throat'' presents the reason of searching light food is user's having sickness, with sore throat as symptom) and change the previous speculation (``I am on a diet, recommend some light but delicious food, like barbecue with less oil'' represents that though user is searching light food, his/her true interest is heavy meals.) 
The procedure is called ``reflection'': an introduced LLM is responsible for deducing underlying information on user's query, reflecting user's characters, current situation, leading to a dynamic and informative user profile for latter usage.
The user profile, together with historical conversations are retrieved by LLM based and embedding based method using query, leading to selected profile and conversation to reduce inference time and generate precise response.
Besides, as previous works are lacking in evaluation for personalization. We proposed a series of evaluation of user profiling, reflection, retrieval and response for measuring their effects to personalization.

The framework aims at knowing the user to provide personalized service (from user perspective). Based on its motivation, it is named as \method{}, which is an Ancient Greek temple of Apollon and well-known for its inscription: ``Know yourself''. The great philosopher, Socrates cited it as his lifelong goal for he thinks knowing oneself is most fundamental and difficult.
We believe that understanding users serves as a very important step in LLM's applications.

In the following, we detail \method{} in Sec.\ref{sec:method}. 
Then in Sec.\ref{sec:experiments}, experiments are demonstrated on user profiling, reflection, retrieval and response.

%% file: content/related-works.tex
Agents (or Autonomous Agents) are defined as a system composed of a series of LLMs for task solving. \cite{wang2023survey} has given comprehensive survey, aiming at providing a unified framework and explanation on LLM agents.
In this section, we have surveyed the related works on dialog agents with personalization. 

\subsection{Dialog Agents}
Dialog agents, or chatbot, has been long studied \cite{mazare2018training}, \cite{serban2017deep} \cite{zhou2020design}, \cite{yu2019gunrock}, \cite{adiwardana2020towards}, \cite{adamopoulou2020overview}.
These early works focus on improving conversation ability to provide human likeness by system design and model training. However, the quality of response is under doubt and they are far from large scale commercial usage.

Recently, the emergence of Generative Pre-trained Transformer (GPT) \cite{radford2018improving} and following chat version (ChatGPT) \cite{brown2020language} announced the era of LLM: most NLP problems including chatbot can be addressed using LLM.
Specially, \cite{lee2023prompted} divided response procedure into multiple modules, solving the difficult task by procedure separation.
\cite{thoppilan2022lamda} pretrained a foundation model on 1.56T words of public dialog data and web text. \cite{madotto2021few} explored prompt-based few-shot learning in dialogue tasks.

These works complete a foundation for dialog system, making it possible to generate reasonable and human-like response with low cost.

\subsection{Personalized Agents}
Personalized agents, agents with personalization or Role-Play \cite{shanahan2023role} denotes endowing agents with characters (real characters such as celebrity or fictitious characters such as cartoon figures, literary characters). This trend of works is often named as ``Digital Character'', aiming at replication of a given person in a digital form. 
\cite{lu2024large} enlightens LLMs’ roleplay capabilities by self-alignment as they have pre-trained on various character profiles and dialogues. 
\cite{shao2023character} teaches LLMs to act as specific celebrity.
\cite{wang2023rolellm} further provides a role-playing framework of data construction and evaluation (RoleBench).

On the contrary, \method{} proposed agents to provide personal service: responses are diverse according to different users. 
In this trend of line, seldom works have done, not even to mention bench-marking the performance of the agents.
\cite{zhang2018personalizing} is an early work to provide consistent personality. It collected data by crowdsourcing and trains models. However, it is a pre-LLM works, mainly focusing on training models for conversation.
\cite{xu2022long} introduced short/long-term memory and conduct dynamic management of persona information. Still, its attention lies at evaluation of dialog such as by perplexity (PPL) \cite{kwon2023and}.
\cite{salemi2023lamp} proposed benchmark for training and evaluating language models for producing personalized outputs.

\subsection{Evaluation of Dialog Agents}
Evaluation of dialog agents has always been a challenge \cite{deriu2021survey}. Unlike CV and NLP problems where labels are well-defined and metrics are formulated, it is difficult to tell whether a response from agent is ``good'' or not.
Before LLM, it is difficult to produce reasonable and meaningful response. Measuring Perplexity (PPL), BLEU, ROUGE becomes a standard method \cite{kwon2023and} \cite{xu2022long}, \cite{liu2016not}, \cite{lowe2017towards}. 
Besides, evaluation is conducted on some specific dialog datasets: \cite{gunasekara2020overview}. \cite{zhang2019dialogpt} conducts a comprehensive evalution on various metrics and datasets.

Apart from objective measurement, dialog agents mostly apply human involved evaluation: They employ human to determine the quality of response. \cite{adiwardana2020towards} use human evaluation and find that the performance is proportional to PPL. 
\cite{lee2023prompted} designed metrics (Sensibleness, Consistency, Interestingness and Preference for pairwise model evalutaion, Sensibleness, Consistency, Engagingness in single model evalution) and provide the user with responses for evaluation.
However, human evaluation introduces great consumption in both cost and time. LLM is becoming a substitute for human evaluator by using a more powerful LLM: \cite{chiang2024chatbot} provided a benchmark platform in a crowdsourced manner. LLMs are competing with each others in multi-turn evaluation. 

However, these evaluations aim at providing reasonable and human likeness response. In \method{}, we pay attention to response with personalization: generation of response closely related to user. 
\cite{salemi2023lamp} proposed 7 tasks for evaluation personalized output and use BM25 as term-matching retrieval model. Similarly, \cite{ao2021pens} presented a personalized headline generation dataset on Microsoft News.

%% file: content/method.tex
\begin{figure}[!htbp]
    \centering
    \includegraphics[width=\textwidth]{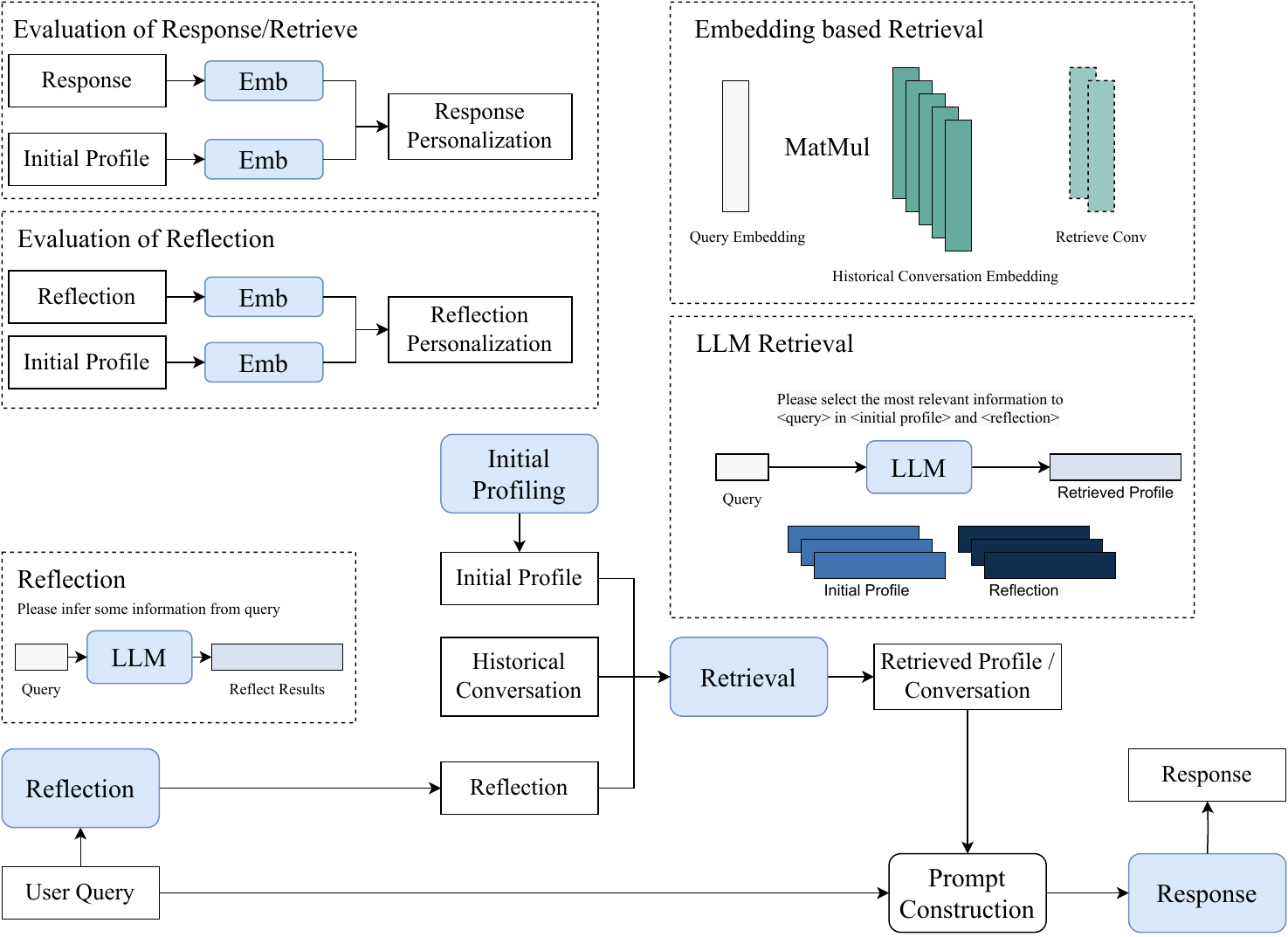}
    \caption{Overflow of \method{}: Rounded box in blue represents modules using LLM or other deep learning models. Transparent Box represents intermediate results from modules. 
    Embedding based and LLM retrieval is demonstrated in upper right corner.
    We provide evaluation protocols for measuring ``Personalization'', as illustrated in left up corner.}
    \label{fig:overflow}
\end{figure}
\method{} is connected by modular prompting of LLMs modules, data structures (for long/short-term storage) and other processing units.
As shown in Fig.\ref{fig:overflow}, \method{} is comprised by \Init{}, \Retrieve{}, \Response{} and \Reflect{} modules:

Consider a specific user. Before serving, in \Init{} module, user's profile is initialized by using corresponding historical interactions, such as purchase records, click sequence and action trace in the Meituan. User profile is represented in text (e.g. ``He is 35-year old, with preference for spicy food''). 


In \Retrieve{} module, query is used to retrieve relevant components in user initial profile $\profile$, reflection $\reflection$ and historical conversations $\conversation$. 
Then the query, together with the retrieved profile and conversation are fed into \Response{} module for utterance generation.

To provide personality, \method{} pays much attention on reflection of user's query and feedback: \Reflect{} module examines user's query to infer information for user understanding. The inferred information is formatted and arranged into user profile as reflection. 

In the following, we will introduce the details of each modules:
\subsection{\Init{}} \label{sec:method:init}
\begin{figure}[!htbp]
    \centering
    \includegraphics[width=\textwidth]{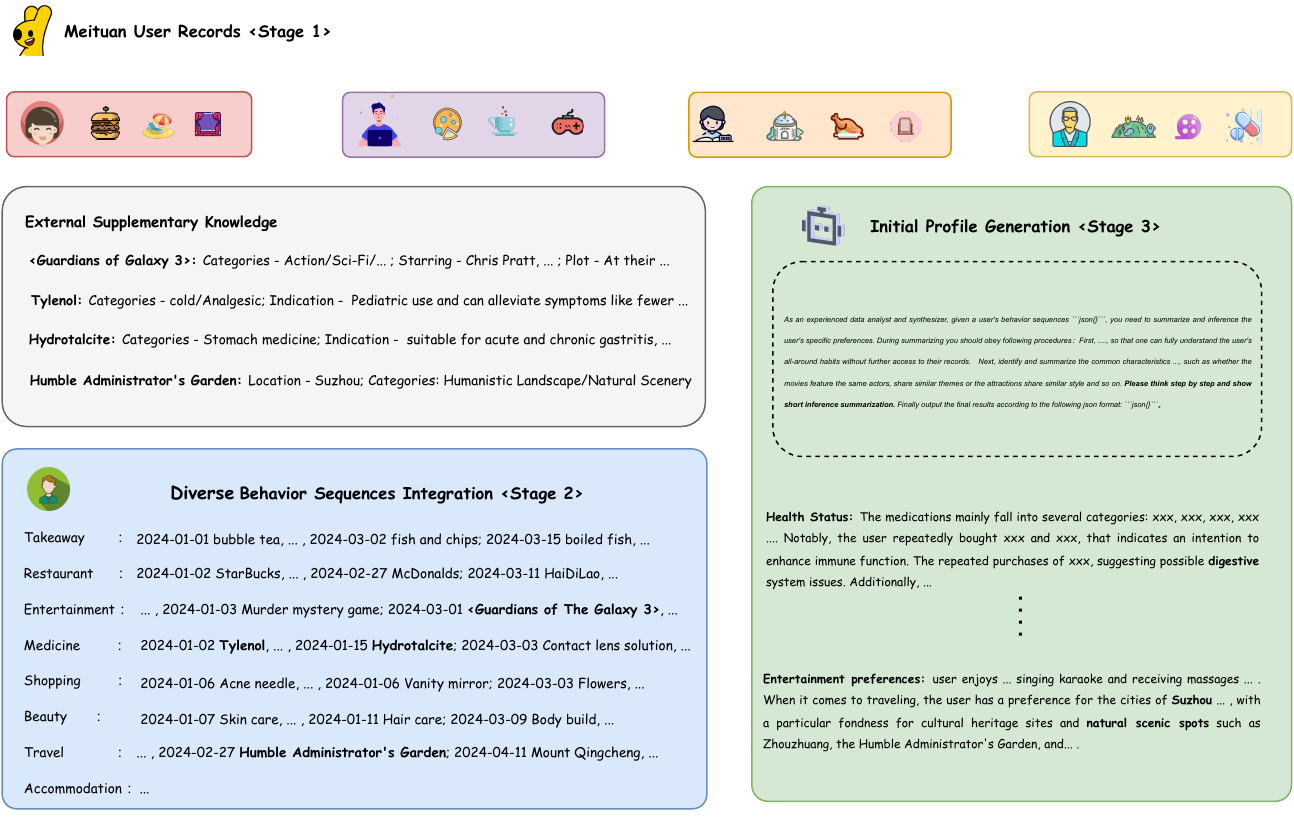}
    \caption{User Profiling Procedures of \method{}: (1) Meituan User Records Collection, (2) Behavior Sequence Integration, (3) User Profile Generation.}
    \label{fig:profiling}
\end{figure}
Departing from traditional approaches, wherein the user profile is typically instantiated based on historical dialogues \cite{zhong2022more},\cite{10.1145/3404835.3462828},\cite{10.1145/3404835.3463239} or rudimentary user information like name, age, location and profession \cite{zhang2018personalizing},\cite{zheng2019pretraining},\cite{ijcai2019p721},\cite{zheng2020personalized}, this module endeavors to construct a comprehensive and informative profile to enhance personalization at the beginning of the dialogue engagement. As illustrated in Figure \ref{fig:profiling}, the user profiling method comprises three stages: (1) Meituan User Information Collection; (2) Preference Sequences Integration; (3) User Profile Generation.

\subsubsection{Meituan User Information Collection}
Composing services ranging from food delivery to movie performances, Meituan's ecosystem enables us to gain a comprehensive understanding of users' diverse preferences in their daily lives. We primarily consider two kinds of information existed in Meituan's transaction system: 1) purchase behavior for its fidelity to the user's genuine intention and 2) item comments, which could help to reveal the user's personality traits and provide a more fine-grained preference signal. 
Additionally, item-level supplementary information such as food categories, movie plots, and medication indications, is introduced to facilitate the induction of user interest or living status. 

\subsubsection{Behavior Sequence Integration}
Despite its thoroughness, the user information is presented in a disorganized manner, which we found to jeopardize the effectiveness of the following preference summarization and personality reasoning methods. Consequently, we integrate the user behaviors and comments into $8$ distinct behavior sequences in time order, according to their source categories and theirs semantic similarities to the behavior categories.

\subsubsection{User Profile Generation}
Based on extracted behavior sequences, we adopt multiple methods to (1) Compress the textual information to ease downstream token limitation; (2) Summarize and reason the user's preference and personality traits, finally leading to user profile. 
The profile is represented in text with $6$ elements: 
\begin{itemize}
    \item Daily Necessities Preference
    \item Entertainment and Travel Preference
    \item User Health Status
    \item Movie and Performances Preference
    \item Diet Preference
    \item Medical and Beauty Preference
\end{itemize}
Profile example are listed in App.\ref{app:response:example1:profile}.
The summary methods could be further categorized into three types: 
1) LLM Generation: Both typical open-source and closed-source LLMs are adopted to generate user interest and personality traits, based on corresponding behavior sequences. 
Prompts for generation is listed in App.\ref{sec:app:init}
2) Rule based: The user profile is represented by frequent statistical information combined with certain recent records; 
3) LLM Compression: Following \cite{jiang2023llmlingua}, the preference sequences are compressed according to the probability distribution over token positions provided by the LLM.


\input{content/retrieve}\label{sec:method:retrieve}

\input{content/reflect}\label{sec:method:reflect}

\subsection{\Response{}}\label{sec:method:response}
After profile, reflection and historical conversations are retrieved, the selected components (denoted with $\tilde{}$ above) are fed into LLM with proper prompt and query:
\begin{eqnarray}
    \response = \text{LLM}(\tilde{\profile}, \tilde{\reflection}, \tilde{\conversation}, \query; \text{Prompt})
\end{eqnarray}
Detailed prompt is listed in App.\ref{sec:app:response:prompt}.

%% file: content/retrieve.tex
\subsection{\Retrieve{}} \label{sec:retrieve}
Long context modeling \cite{dong2023survey} and Retrieval Augmentation Generation (RAG) \cite{zhu2023large} answer to the question of how LLM deals with long input. 
Long context modeling replies on pretraining LLM tailored for long inputs while RAG utilizes external memory and information abstraction before feeding input with normal length into LLM.

\method{}'s underlying philosophy lies in solving a complex personal response task by separation into various simple sub-tasks. Before feeding input for response generation, \Retrieve{} is designed to retrieves most relevant profiles and historical conversation according to user's query.

\method{} sets up $3$ components for retrieval: 
(1) Initial Profile: user description based on interaction history in Meituan. (Sec.\ref{sec:method:init}).
(2) Reflection: user description summarized during conversation (Sec.\ref{sec:method:reflect}).
(3) Historical Conversation.
For each component, \Retrieve{} implements 3 types of retrieval methods.

\noindent \textbf{Full Content}: Practically, a well-trained foundation model is able to retrieve relevant information according to query to generate proper response. \Retrieve{} uses full content as a baseline.

\noindent \textbf{Embedding based Retrieval}: Retrieved components are constituted by a list of elements. 
Specially, in Initial Profile, $6$ elements are derived to represent user's corresponding respects.
Reflection progressively collect user profile during conversation. Each query generates a reflection, which is denoted as an elements.
Each conversation (including query and response) is regarded as en elements.

Each elements are embedded by Text2Vec \cite{Text2vec}. During retrieval, the embedded query is used to retrieve the top k closest elements in components, leading to the retrieved elements for latter response. 
For example, retrieved profile $\tilde{\profile}$ is formulated as:
\begin{eqnarray}\label{eq:embedding-based-retrieve}
    \tilde{\profile} = \text{TopK}(\text{emb}(\profile), \text{emb}(\query), k)
\end{eqnarray}
Correspondingly, historical conversation is retrieved:
\begin{eqnarray}\label{eq:embedding-based-retrieve:conversation}
    \tilde{\conversation} = \text{TopK}(\text{emb}(\conversation), \text{emb}(\query), k)
\end{eqnarray}
Normally, we set $k=1$.

\noindent \textbf{LLM based Retrieval}: Instead of retrieving single element from components, LLM based method uses all contents from the components and generates relevant information by leveraging LLM. 
Formally:
\begin{eqnarray}\label{eq:llm-based-retrieve}
    \tilde{\profile} = \text{LLM}(\profile, \text{Prompt})
\end{eqnarray}
Prompt for retrieval is listed in App.\ref{sec:app:retrieve}.

Though 3 retrieval methods are provided, components are not suitable for all retrieval methods. For example, as conversation progresses, length of conversation content explodes and can not be fed in LLM retrieval, nor as full content for latter response.
\method{} uses the retrieval methods for each component as shown in Table \ref{tab:retrieve:method-applied-component}.
\begin{table}[]
    \centering
    \begin{tabular}{c|c|c}
    \hline \hline
    Component & Embedding-based & LLM-based  \\ \hline
    Initial Profile & $\times$ & $\checkmark$ \\ \hline
    Reflection & $\checkmark$ & $\checkmark$ \\ \hline
    Historical Conversation & $\checkmark$ & $\times$ \\ \hline
    \end{tabular}
    \caption{\method{} uses combination of embedding/LLM-based retrieval methods for retrieved component.}
    \label{tab:retrieve:method-applied-component}
\end{table}

%% file: content/reflect.tex
\subsection{\Reflect{}}\label{sec:method:reflect}
\Reflect{} module aims at maintaining a precise and update-to-date description of user, providing foundation for personalized conversation. 
Specially, \Reflect{} infers information as reflection $\reflection$ from user's query $\query$ and its feedback action (if applicable).
It is formulated as:
\begin{eqnarray}\label{eq:reflect:reason}
    \reflection = \Reflect{}(\query)
\end{eqnarray}
$\reflection$ describes reflection of user, depending on the query. For example, given user's query: \emph{I need some advice of Beijing cuisine in Wangjing District}, \Reflect{} is able to infer that user has interest in \underline{Beijing cuisine} and his possible working or living area is \underline{Wangjing District}, which leads to:
\begin{eqnarray}
    \reflection = \text{Diet preference: Beijing cuisine. Possible working/living area: Wangjing District}
\end{eqnarray}

The resulting reflection $\reflection$ serves as supplement for total user profile $\profile$. During serving, $\reflection$ and $\profile$ are retrieved respectively according to query for precise response.



 

%% file: content/experiments.tex
Evaluation of dialog agents has always been a challenge. 
Tradition topics in machine learning such as Computer Vision (CV), Natural Language Processing (NLP) conduct evaluation on well-define setting, where labels are given without ambiguity, such as image classification, objection detection and named entity recognition. 
Performance of the proposed method can be directly measured with input features (images, sentences and etc.) for prediction and labels for error computation.
However, dialog agents aim at ``How intelligent / humanoid / satisfying it is''. Given a user query, there are abundant of responses that can be regarded as satisfactory. Though human intervention is able to discriminate the difference and score the response, it is still difficult and consuming for scalable and quantification evaluation.
Basically, it measures to what extend the dialog agent performs, instead of a specific output (such as logits for deep learning models) for loss calculation. 

\begin{figure}[!htbp]
    \centering
    \includegraphics[width=\textwidth]{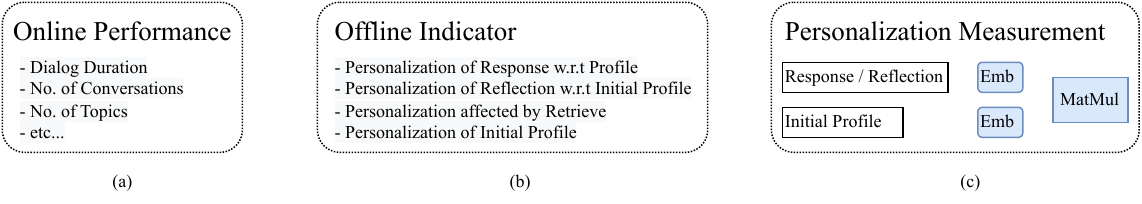}
    \caption{Experiments workflow of \method{}. 
    Figure (a): Metrics of online performance. User experiences are reflected in dialog duration and etc. This metric is adopted after official launch and commonly used in commercial dialog agents. 
    Figure (b): Offline indicator, especially ``Personalization'' focused in \method{}. Performance of each modules are measured. 
    Figure (c): Qualification of Personalization given profile and corresponding response. It is used as a base measurement for module evaluation in Figure (b).}
    \label{fig:exp:experiment-worflow}
\end{figure}

Based on that, commercial dialog agents use online performance such as dialog duration, number of conversations for evaluation and etc (Fig.\ref{fig:exp:experiment-worflow}(a)). A higher value represents users are tending to converse and ask assistance from the agents, which indirectly reflects a good performance of the dialog agent. 
However, the online statistic are collected in official deployment, which is a very last step in development. Besides, it is difficult to apply online data directly in algorithm design, intermediate offline indicators are necessary.

Traditionally, manual / GPT-based evaluation is widely used in offline response evaluation. The setting employs human or human-like GPT models to examine responses and provide scores according to different standards, such as ``fidelty'' (how the response resembles human's wordings), ``truthful'' (whether the response is faking information) and etc. Though it is straightforward and directly corresponding to final online performance, it is time/cost-consuming and it relies on subjective judgement who may also have bias.
On the contrary, automatic evaluation is commonly used in specific machine learning problems such as image classification, it is rare that dialog agents perform measurable and specific tasks.

In this work, we proposed a series of evaluation protocols for dialog agents, especially on their ability to provide personalization service. 
Traditionally, evaluation of LLM based dialog agents heavily reply on a third-party judge LLM, normally more powerful but more expensive. 
Besides, it is very difficult to perform scalable evaluation due to the limited input length for the judge LLM.
In stead, the evaluations proposed by \method{} reply on automatic evaluation for quick and cheap evaluation.
As shown in Fig.\ref{fig:exp:experiment-worflow} (b), we design several offline indicators to measure the personalization ability of each module in dialog agents. These indicators serve as an intermediate step for online performance (Fig.\ref{fig:exp:experiment-worflow} (a)) and measurable metric during offline optimization. 
These offline indicators are quantifiable and scalable, whose central measurement replies on personalization measurement in Fig.\ref{fig:exp:experiment-worflow} (c): Personalization of dialog agents is computed by embedding response / reflection / initial profile and matrix multiplication for similarity between them.
\method{}'s personalization ability is fulfilled in modules which are connected to provide a final implementation. 

\begin{figure}[!htbp]
    \centering
    \includegraphics[width=\textwidth]{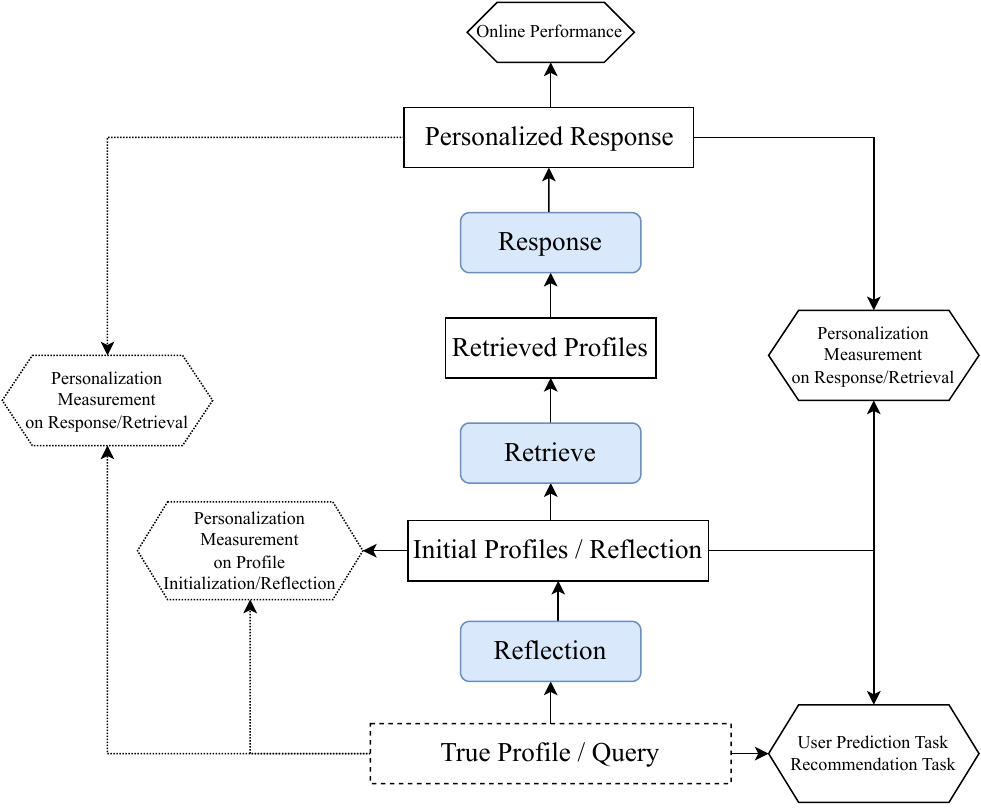}
    \caption{Ideal design v.s. Practical situation in personalization measurement. Dashed boxes (left side) represents ideal evaluation but infeasible due to inaccessible of true profile. Solid boxes (right side) illustrates the practical solution: Performance of \Response{} and \Retrieve{} is measured by personalization between response and initial profile. Performance of profile initialing is evaluated on User Prediction Task and Recommendation Task.}
    \label{fig:exp:experiment-idea}
\end{figure}
We examine the performance of every module and design its corresponding metrics. The overall idea of personalization evaluation in \method{} lies in: \emph{whether the proposed module contributes to the distinguishability of response w.r.t user}.

Ideally, as shown in Fig.\ref{fig:exp:experiment-idea}: user's true profile is accessible. \Init{} summarizes the true profile into initial profile. \Retrieve{} selects relevant information of initial profile \footnote{We omit the retrieval for historical conversation for clear description.} for \Response{}. \Reflect{} updates the profile by inferring user's query, leading to reflection. 
Similarity (or personalization) of initial profile / response / reflection w.r.t true profile can be measured by certain metric.
However, in practice, true profile is inaccessible. Besides, it is difficult to define similarity metric. 
In \method{}, we address the first question by relaxing the similarity of response / reflection w.r.t \textbf{initial profile}, and leaving initial profile evaluation as another question.
The second challenge is tackled by defining similarity metric in Fig.\ref{fig:exp:experiment-worflow} (c).
Specially, we first define similarity metric as matrix multiplication between embedding, as mentioned in Sec.\ref{sec:experiments:personalized-response}. 
This definition enables quantifiable and scalable measurement. It is further utilized as basic measurement in other modules' evaluation.
In \Retrieve{}, we measure whether the retrieved profiles and historical conversations are relevant to user's query. We use the similarity between response using retrieved profiles w.r.t initial profile as measurement.
\Reflect{} updates user's profile as conversation proceeds. 
It's measurement is conducted by personalization between reflection w.r.t. initial profile. Besides, as conversation progresses, the similarity is supposed to increase.
\Response{} module aggregates information from previous modules for final utterance generation, we measure its personalization between response w.r.t initial profile. 
Finally, for evaluation of \Init{}, we evaluate user profile initialization on how it contribute to personalized response and user distinguishability.

In the following, our experiment section begins by introduction of datasets in Sec.\ref{sec:experiments:dataset}, which is a real-world user behavior records collected from Meituan Delivery. 
Then we introduce the quantified measurement of personalized response w.r.t profile in Sec.\ref{sec:experiments:personalized-response}. 
In the following parts, modules' experiments are illustrated.

\subsection{Dataset} \label{sec:experiments:dataset}
As shown in Fig.\ref{fig:profiling} (Stage 1 \& 2), we construct a dataset based on users' transaction records in Meituan platform. 
Specially, we first aggregate the interaction records for distinctive user. Then the records are integrated into six temporal behavior sequences.
Sensitive information in privacy security concern are desensitized: 1) The users are selected under a strategy balancing randomness and level of engagement.
2) The length of sequence is balanced by randomly selecting parts of records for highly engaged users.
3) Certain proper nounes such as brand name, location are converted to some unified representations.
This dataset provides comprehensive and realistic description of users' life trajectory, including dining, shopping, entertainment, medicine, beauty, travel, accommodation and etc.

We split each user's behavior sequences into training and testing set in time order. To guarantee the quality of the behavior sequence, the users with overall training sequence less than 100 items and testing sequence less than 60 items are filtered. 
The basic statistics of the dataset is provided in Table \ref{tab:statistics}, where the overall number of users in each behavior sequence is noted as $user\_nums$. $item\_nums\_avg$ and $total\_item$ refer to the average and total number of items in each kind of behavior sequence respectively.  
$tokens$ represents the average approximated number of tokens that the raw behavior sequence occupies, which is calculated using OpenAI's tiktoken tool.

\renewcommand{\arraystretch}{2}
\begin{table}[]
\scalebox{0.525}{
\begin{tabular}{|c|c|c|c|c|c|c|}
\hline
\cellcolor[HTML]{C0C0C0} & \cellcolor[HTML]{C0C0C0} & \cellcolor[HTML]{C0C0C0} & \cellcolor[HTML]{C0C0C0} & \cellcolor[HTML]{C0C0C0} & \cellcolor[HTML]{C0C0C0} & \cellcolor[HTML]{C0C0C0} \\
\multirow{-2}{*}{\cellcolor[HTML]{C0C0C0}\textbf{\begin{tabular}[c]{@{}c@{}}Behavior Categories /\\ Statistics\end{tabular}}} & \multirow{-2}{*}{\cellcolor[HTML]{C0C0C0}\textbf{Diet}} & \multirow{-2}{*}{\cellcolor[HTML]{C0C0C0}\textbf{Daily Shopping}} & \multirow{-2}{*}{\cellcolor[HTML]{C0C0C0}\textbf{Medicine}} & \multirow{-2}{*}{\cellcolor[HTML]{C0C0C0}\textbf{Movie Performance}} & \multirow{-2}{*}{\cellcolor[HTML]{C0C0C0}\textbf{Beauty}} & \multirow{-2}{*}{\cellcolor[HTML]{C0C0C0}\textbf{\begin{tabular}[c]{@{}c@{}}Entertainment \&\\ Accommodation \end{tabular}}} \\ \hline
user\_nums & 432 $|$ 432 & 426 $|$ 430 & 393 $|$ 410 & 252 $|$ 332 & 353 $|$ 392 & 418 $|$ 426 \\ \hline
item\_nums\_avg & 39.09 $|$ 30.29 & 26.16 $|$ 23.47 & 14.68 $|$ 10.03 & 3.23 $|$ 2.42 & 4.56 $|$ 3.03 & 10.68 $|$ 8.22 \\ \hline
total\_items & 10611 $|$ 8503 & 9111 $|$ 8241 & 2774 $|$ 2117 & 115 $|$ 109 & 43 $|$ 45 & 3705 $|$ 2843 \\ \hline
tokens & 1174.98 $|$ 908.62 & 1279.91 $|$ 1154.42 & 745.82 $|$ 511.76 & 227.46 $|$ 160.03 & 50.87 $|$ 35.04 & 486.14 $|$ 375.29 \\ \hline
\end{tabular}
}
\caption{Basic Statistics of Dataset. The dataset covers users interactions over a five-month period. Interaction records from the first three months serve as the training sequence, while the remaining two months are used for testing. Statistics of train set and test set are separated by the symbol $|$ in each column.}
\label{tab:statistics}
\end{table}

\subsection{Evaluation of \Init{}} \label{sec:experiments:init}
In this section, we first introduce the principles that a user profile with good quality should adhere to. Then the proposed automatic evaluation metrics for direct user profile quality evaluation are illustrated, followed by the detailed experiment setting and corresponding experimental results. 

\subsubsection{Quality Principles}
Specifically, a user profile with good quality should follow:
(1) \textbf{Distinctiveness}: user profile should compose individual traits and preferences that set users different, rather than standardized phrases. It is supposed to be distinctive enough to enable personalized response generation of dialog agent; 
(2) \textbf{Truthfulness}: The user profile should be faithful to origin user behaviors and comments content without severe hallucination. It should be able to reflect users' genuine interests or point and thus be beneficial to the honesty of dialog agent; 
(3) \textbf{Inductiveness}: Instead of enumerating raw data, the user profile should summarize and reason user's preferences and characteristics. Besides of decreasing the number of token used in downstream tasks, an inductive profile can generalize to future behavior and thus be beneficial to the long-term helpfulness of the dialog agent. 

\subsubsection{Evaluation Metrics}
\begin{figure}[!htbp]
    \centering
    \includegraphics[width=\textwidth]{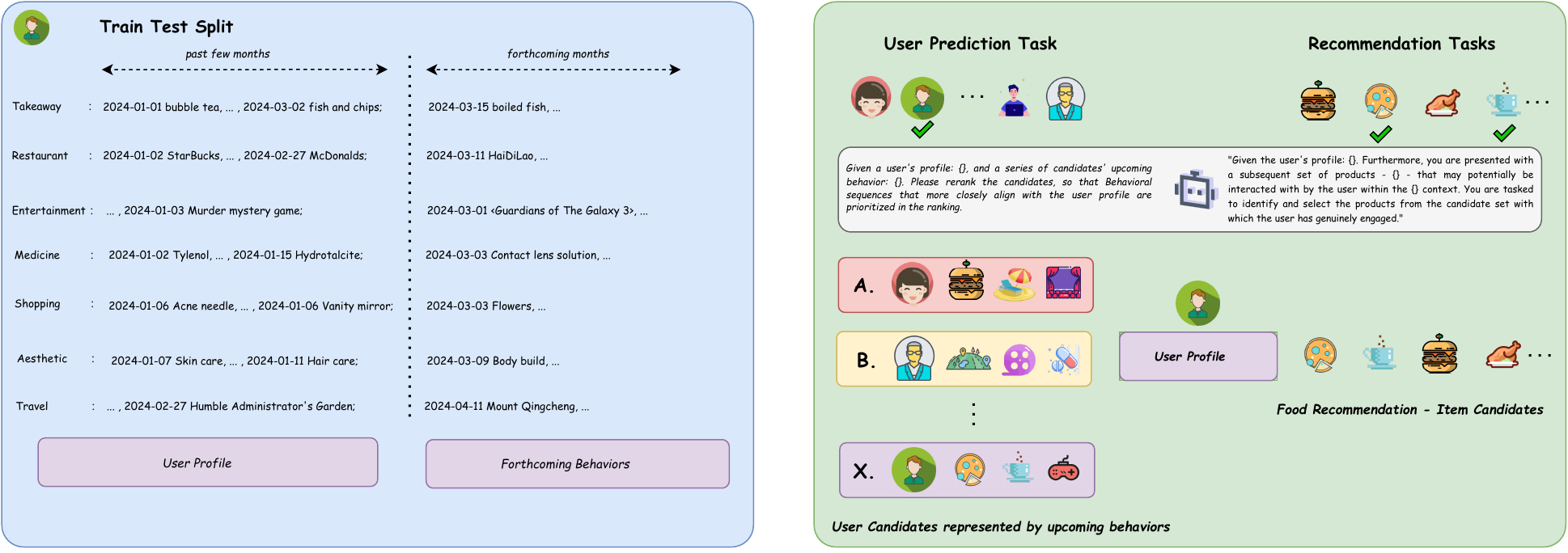}
    \caption{Illustration Evaluation of \Init{}: 
    In the left figure, we first collection user behavior sequence and split it into training/test set, according to a specific period. 
    In right part, we illustrate the user prediction task and recommendation task. User prediction task aims at discriminating ground truth user from a group of candidates given the corresponding profile. Recommendation task aims at selecting most relevant items for a user given corresponding profile.}
    \label{fig:quality}
\end{figure}

Following the profile principles, we propose two tasks for comprehensive user profile evaluation: user prediction task and recommendation task, which are illustrated in Figure \ref{fig:quality}.
\begin{itemize}
\item \textbf{User Prediction Task}: 
For a specific user $U_g$ with profile $\profile_g$ (generated by training set), given a group of candidate users composing ground truth user $U_g$ and $k$ negative users $<U_g, U_1, ..., U_k>$. 
The users are represented by their behavior sequences $<\Tilde{B}_{g}, \Tilde{B}_{1}, ..., \Tilde{B}_{k}>$ in test set, the task aims at discriminate $U_g$ from the candidates based on the given profile. Mathematically, the task can be formulated as:
\begin{eqnarray}\label{eq:user-recommendation}
    \text{Prob}(U_g|\profile_g, <\Tilde{B}_{g}, \Tilde{B}_{1}, ..., \Tilde{B}_{k}>)
\end{eqnarray}
In Eq.\ref{eq:user-recommendation}, we measure the probability of finding the ground truth user $U_g$ from $<U_g, U_1, ..., U_k>$ (represented by $<\Tilde{B}_{g}, \Tilde{B}_{1}, ..., \Tilde{B}_{k}>$) given $\profile_g$.
By the definition, the user profile should adhere to the three principles, especially distinctiveness, to achieve success on this task. 
Specifically the user profile should be distinct and faithful enough to differentiate from negative samples, and be inductive to align with forthcoming behavior sequence $\Tilde{B}_{g}$. 
Eq.\ref{eq:user-recommendation} generates the probability for all candidate users, $Acc@1$ is denoted as the ground truth user with highest probability.

\item \textbf{Recommendation Task}:  Given $\profile_g$ and a candidate item set that includes n positive items $<I_{\text{pos}_i},..., I_{\text{pos}_n}>$ sampled from $\Tilde{B_g}$ and k negative items $<I_{\text{neg}_1},..., I_{\text{neg}_k}>$ sampled from item pool, the recommendation task aims at prioritizing positive samples over negative ones, by leveraging user profile. Mathematically, the task can be formulated as:
  \begin{eqnarray}\label{eq:item-recommendation}
      \text{Prob}(<I_{\text{pos}_1},..., I_{\text{pos}_n}> \xrightarrow{} <I_{\text{neg}_1},..., I_{\text{neg}_k}>|\profile_g, <I_{\text{pos}_1}, I_{\text{neg}_1},...,I_{\text{pos}_n},I_{\text{neg}_n}>)
  \end{eqnarray}
  Given a profile $\profile_g$, Eq.\ref{eq:item-recommendation} measures the probability of ranking the positive items (belongs to $U_g$) above the negative items. $\alpha \xrightarrow{} \beta$ stands for $\alpha$ prioritizes $\beta$ in an ordered set.
  Referring to the task specification, a faithful and inductive profile is necessary for predicting forthcoming interacted items and achieving success on the task. Besides, intuitively better performance on the task is beneficial to the personalization of downstream dialog agent, especially in recommendation related topics. 
  We report $NDCG@5$ and $Recall@5$ that positive items are ranked within the top 5.
\end{itemize}


\subsubsection{Experiment Setting}
To the concern of tokens limitation, disturbance of long-range dependency capacity,  and evaluation cost, we set the number of negative samples $k$ to $4$ and $7$ in user prediction task and recommendation task respectively. And the number of positive items $n$ in recommendation task is set to $3$. 
So that the user prediction task could be seen as a Multiple Choice Question task with each question paired with $5$ long choices and the recommendation task as a Multiple Answer Question task with each question paired with $10$ choices and $3$ answers inside. 
In item recommendation, we perform evaluations by selecting items from different behavior sequences: diet, daily shopping, health and overall. 

For evaluation, GPT-4 is adopted for its well balance between task performance and inference cost. 
Except LLMs generation method, we consider two baselines for profile initialization: (1) Rule based: Where the user's $6$ preferences are represented by the high-frequency statistical information extracted from corresponding behavior sequences; (2) LLM Compression: Following \cite{jiang2023llmlingua}, the user's $6$ preferences are the compression result of the corresponding origin behavior sequences, according to the probability distribution over token positions provided by the LLM. 
The profiles generated by the different methods are sampled in App.\ref{sec:app:init}. 
Considering the downstream token budget, during profile generation, we manually adjust hyperparameters of each method, such as prompt in LLM generation methods and the number of statistical quantities in rule based method, to guarantee each part of the profile converge to 300 tokens and the whole profile is around 1600 tokens. 

Furthermore, for fair comparison the length normalized score is reported, labeled as length normalization (LN) in Table \ref{tab:quality}, where the score penalty is conducted when the profile length exceeds the threshold that $300$ for recommendation tasks and $1600$ for user prediction task. 
Specially, for a score $s$ with profile length $n$, score penalty is conducted:
\begin{eqnarray}
    s = \left\{
    \begin{aligned}
        s \quad & \text{if} & n \leq \text{LN} \\
        \frac{s}{n / \text{LN}}  \quad & \text{if} & n > \text{LN}
    \end{aligned}
    \right.
\end{eqnarray}
Besides, we report the compression ratio as the percentage of profile's token number over behavior sequence's token number. A small compression ratio leads to more compact and succinct profile.

\subsubsection{Experimental Results}
The experimental results are presented in Table \ref{tab:quality}. 
Surprisingly, Rule based methods overpass LLMLingua on all metrics, which may due to the fact that the compression result of LLMLingua tend to retain original behavior sequence, especifically when it is short, and thus in short of 
inductiveness to generalize to test sequences. 
Attribute to better exploit the rich textual information in the behavior sequences and comments, LLM generation methods lead to more inductive and discriminative profiles, and thus show great superiority over other methods on both tasks. As expected, among LLM generation methods, GPT-4 performs best for producing less hallucination and better balance the detail and induction under the token limitation.

Besides, GPT-4 generates the smallest compression ratio over other methods, showing that it is able to summarize the content of original behavior sequence into a succinct and representative profile.

\renewcommand{\arraystretch}{2}
\begin{table}[!htbp]
\scalebox{0.525}{
\begin{tabular}{|c|cclclcl|cl|c|}
\hline
\rowcolor[HTML]{C0C0C0} 
\textbf{Model Name} & \multicolumn{7}{c|}{\cellcolor[HTML]{C0C0C0}\textbf{Recommendation Tasks}}                                                                                                                      & \multicolumn{2}{c|}{\cellcolor[HTML]{C0C0C0}\textbf{User Prediction Task}} & \textbf{Compression Ratio} \\ \hline
                    & \multicolumn{1}{c|}{OVERALL}                   & \multicolumn{2}{c|}{Diet}                     & \multicolumn{2}{c|}{Daily Shopping}           & \multicolumn{2}{c|}{Health}                   & & &                            \\ \cline{2-10}
\multirow{-2}{*}{/} & \multicolumn{7}{c|}{Recall@5 $|$ NDCG@5(LN-300)}                                                                                                                                                  & \multicolumn{2}{c|}{Acc@1(LN-1600)}                                       & \multirow{-2}{*}{/}        \\ \hline
GPT-4               & \multicolumn{1}{c|}{\textbf{0.5776 $|$ 0.5519}}   & \multicolumn{2}{c|}{0.4939 $|$ 0.4779}          & \multicolumn{2}{c|}{\textbf{0.5914 $|$ 0.5644}} & \multicolumn{2}{c|}{\textbf{0.6476 $|$ 0.6133}} & \multicolumn{2}{c|}{\textbf{0.7513}}                                       & \textbf{0.3987}            \\ \hline
Baichuan-13B        & \multicolumn{1}{c|}{{0.5434 $|$ 0.5171}}      & \multicolumn{2}{c|}{{0.5098 $|$ 0.4796}}    & \multicolumn{2}{c|}{{0.5382 $|$ 0.5123}}    & \multicolumn{2}{c|}{0.5823 $|$ 0.5596}          & \multicolumn{2}{c|}{0.6166}                                                & {0.4202}               \\ \hline
ChatGLM3-6B         & \multicolumn{1}{c|}{0.5243 $|$ 0.5080}            & \multicolumn{2}{c|}{\textbf{0.5674 $|$ 0.5207}} & \multicolumn{2}{c|}{0.4603 $|$ 0.4860}          & \multicolumn{2}{c|}{0.5451 $|$ 0.5174}          & \multicolumn{2}{c|}{{0.6405}}                                          & 0.4270                     \\ \hline
Rule based          & \multicolumn{1}{c|}{0.5151 $|$ 0.4880}            & \multicolumn{2}{c|}{0.4414 $|$ 0.4149}          & \multicolumn{2}{c|}{0.5077 $|$ 0.4799}          & \multicolumn{2}{c|}{{0.5963 $|$ 0.5692}}    & \multicolumn{2}{c|}{0.5069}                                                & 0.4934                     \\ \hline
LLMLingua           & \multicolumn{1}{c|}{0.5004 $|$ 0.4813}            & \multicolumn{2}{c|}{0.4751 $|$ 0.4576}          & \multicolumn{2}{c|}{0.4625 $|$ 0.4400}          & \multicolumn{2}{c|}{0.5635 $|$ 0.5464}          & \multicolumn{2}{c|}{0.4620}                                                & 0.4331                     \\ \hline
\end{tabular}}
\caption{Evaluation of \Init{}: Performance of Recommendation Tasks and User Prediction Task is reported. 
In item recommendation, we report the Recall@5 and NDCG@5 under LN=300, meaning that we calculate the recall / NDCG metric for ground truth items in top 5 ranked results, the metric is penalized if profile's length exceeds 300. 
Similarly, we measure whether the ground truth user is the predicted user. The accuracy score is penalized if profile's length is over 1600.
Compression ratio represents the ratio between the number of tokens in profile generated by different methods and the token number of origin behavior sequence. Lower the ratio leads to more succinct profile. 
}
\label{tab:quality}
\end{table}

\subsection{Evaluation of \Response{}} \label{sec:experiments:personalized-response}
\method{} aims at providing personalized response: users with different profile are expected to be served with diverse under a same query. 
For example, a user with interest in outdoor activities should be recommend with high-protein food (instead of light food, such as salad) under the query ``Help me with a menu for this lunch".

However, it is difficult to determine whether a response is personalized: 
1) There is no prior works on personalization measurement. 
2) Labeling personalization by human effort is consuming and biased. 
3) Given a query and profile, personalization measurement of corresponding response by LLM can be regarded as ability of instruction following. The difference between ability for personalization and instruction following should be studied.

Based on the above-mentioned concerns, we proposed a metric for response personalization given query and profile. Formally, for a dialog system $\diag$ parameterized by $\theta$, where $\theta$ includes LLM's parameters, prompts and hyperparameters. 
$\diag$ receives query $\query$ and profile $\profile$ as input and generate response $\response = \diag(\query, \profile)$ as output. A metric for personalization $\PersonalizedMetric(\profile, \response)$ is defined with the following properties:
\begin{theorem}[Properties of Personalization]
Given different profile $\profile_i, \profile_j$ and corresponding $\response_i$, $\response_j$, personalization between a pair of profile and corresponding response should be larger than that of the profile and other response:
\begin{eqnarray} \label{eq:experiments:personalization-inequality}
    \PersonalizedMetric(\profile_i, \response_i) \quad \text{or} \quad \PersonalizedMetric(\profile_j, \response_j) > \PersonalizedMetric(\profile_j, \response_i) \quad \text{or} \quad \PersonalizedMetric(\profile_i, \response_j) 
\end{eqnarray}
\end{theorem}
We further formulate $\PersonalizedMetric$ as:
\begin{eqnarray}\label{eq:experiments:personalization}
    \PersonalizedMetric(\response, \profile) = \text{embedding}(\response)^\top \text{embedding}(\profile)
\end{eqnarray}
Basically, Eq.\ref{eq:experiments:personalization} simply embeds $\response$ and $\profile$ to a same space and calculate their similarity using inner production. In practice, we use Text2Vec \cite{Text2vec} for embedding.

Though simple and straight forward, Eq.\ref{eq:experiments:personalization} is proven to be a practical tool for measuring the instruction following ability of $\diag$. 
Though repeat of profile as response is able to achieve highest personalization score, LLM's completion ability ensures the response to be reasonable and normal.

Besides, personalization of $\response$ is highly related to query: dialog system is expected to provide personalized response for subjective query, such as activity arrangement, food recommendation. But objective query for common knowledge assistance requires response without injection of profile.

Therefore, to evaluate the usage of Eq.\ref{eq:experiments:personalization} and examine whether it can be used in personalization measure, we prepare a series of subjective queries and user profiles. We use a simple $\diag$ (simply feeding profile and query to LLM) to generate response and calculate the $\PersonalizedMetric$. Full prompt incorporating profile and query and subjective queries are listed in App.\ref{sec:app:response:prompt}.

Specially, we select 11 users and generate 5 personal questions. GPT-4, GPT-3.5, Baichuan-2 are used as foundation model to generate response. 
For each query, we use all profiles provided in Sec.\ref{sec:experiments:init} to generate corresponding responses and calculate $\PersonalizedMetric$. 
We further compute $\PersonalizedMetric$ across user and response, leading to a 2-dimension heat map where elements with index $(i,j)$ representing 
$\PersonalizedMetric(\profile_i, \response_j)$. 
By retrieving the top k $\PersonalizedMetric$ in $l$-th row and determining whether $l$-th response lies in the retrieval set, we further derive a metric for personalization matching accuracy among the profile set. 
The accuracy represents the similarity between the generated response and corresponding initial profile, among all user candidates.

\begin{table}[!htbp] 
    \centering
    \begin{tabular}{c|c|c}
        \hline \hline
        Model name & Acc (Top-1) (\%) & Acc (Top-3)(\%) \\ \hline
        Baichuan2-Turbo & 10.1 & 29.0  \\ \hline
        GPT-3.5 & 21.8 & 43.6 \\ \hline
        GPT-4 & 25.5 & 41.8 \\ \hline
    \end{tabular}
    \caption{Peronalization of response w.r.t initial profile. Accuracy is calculated by determining whether the $\PersonalizedMetric$ of response and corresponding initial profile lies in the top $k$ $\PersonalizedMetric$ of response w.r.t to all initial profile candidates.}
    \label{table:response}
\end{table}

The experiment can be considered a 11-class classification for 5 times, the average topk accuracy is reported to measure the perosonalization of response.
As shown in Table \ref{table:response}, Baichuan2 achieves a 10.1\% accuracy among 11 users, which is rather a guess for each query. It shows that Baichuan2 as response model is hardly able to generate personal response.
GPT-3.5 and GPT-4 are able to reach a considerable performance. Though GPT-4 achieves better performance in Top1, GPT-3 outperforms a margin in Top3.

The detailed prompt, initial profile and response is listed in App.\ref{sec:app:response}.

\subsection{Evaluation of \Retrieve{}}
\Retrieve{} aims at retrieving most relevant profiles $\selectedProfile$ for query $\query$ to generate personalized response $\response$. 
The contribution of retrieval to response generation lies at: Whether retrieval select the most relevant information by ignoring the misleading information.
Though powerful foundation model is able to select relevant information, input profile with short length reduces inference time and consumption.

Based on the consideration, the evaluation the retrieval follow the idea that: a more precise retrieval leads to more personalized response. Similarly as Sec.\ref{sec:experiments:personalized-response}, we measure the similarity between initial profile and response generated by different retrieval methods.
Though \Retrieve{} takes place in both conversation history and profile (initial profile and reflection), we only evaluate retrieval on initial profile for simplicity and without loss of generalization.

\Retrieve{} is implemented by using different methods: 
We use a full-profile method as baseline. As mentioned in Sec.\ref{sec:retrieve}, we compare embedding based retrieval and LLM based retrieval, and their combination as multi-path retrieval. 
GPT-3.5 is utilized as the response module.
We also report the token number of retrieved profiles under retrieval methods.
\begin{table}[!htbp] 
    \centering
    \begin{tabular}{c|c|c|c}
        \hline \hline
        Method & Acc (Top-1) (\%) & Acc (Top-3)(\%) & Profile Tokens\\ \hline
        Full Profile & 21.8 & 43.6 & 1205.5 \\ \hline
        Embedding based & 20.0 & 32.7 & 255.8\\ \hline
        LLM based & 29.1 & 52.7 & 164.1\\ \hline 
        Embedding+LLM & 16.3 & 36.3 & 420.5 \\ \hline
    \end{tabular}
    \caption{Peronalization of response under different retrieval methods. Top1/3 accuracy and length of retrieved profile are compared.}
    \label{table:retrieve}
\end{table}

Table \ref{table:retrieve} demonstrates performance of various retrieval methods: 
Full profile baseline achieve a considerable performance, showing that LLM is able to select relevant information by feeding all profiles.
Embedding based retrieval method shows the worst performance, for it may fail to retrieve the corresponding profile. Besides, important information may reside in two different profiles, leading to incomplete retrieval.
LLM based method achieves the best performance while using the fewest profiles. 
It shows that LLM is able to abstract important information. 

LLM based retrieval can be considered as a pre-task before a main LLM response by elimination processing complexity for the main task. In latter development, a cheaper foundation model can be utilized for LLM based retrieval.
Finally, combination of embedding and LLM based retrieval method lead to bad performance, showing that retrieving irrelevant information severely harm performance in personalization.

\subsection{Evaluation of \Reflect{}}
\begin{figure}[!htbp]
    \centering
    \includegraphics[width=\textwidth]{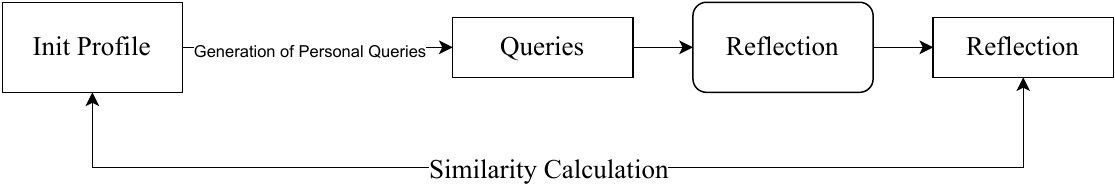}
    \caption{Experiments workflow of \Reflect{}.}
    \label{fig:exp:experiment-reflection-worflow}
\end{figure}
\Reflect{} updates user profiles. A better update policy captures user's information exposed during conversation and generates more personalized response. 
Basically, it can be regarded as an improved version of single-turn personalized conversation: As more and more information is absorbed, performance of \method{} is improved.
A natural idea is to collect the variation of response personalization in multi-turn conversation to demonstrate the performance of \Reflect{}.

However, personalization of final response is effected by various modules. To sophisticatedly examine the performance, we have designed the following experiments protocol:
As shown in Fig.\ref{fig:exp:experiment-reflection-worflow}: For each user, we used his initial profile (from Sec.\ref{sec:method:init}) to generate $K$ personal queries by GPT-3.5-Turbo. 
The queries reflect curtain properties of the user. For example, a user with interest in spicy cuisine is tending to query \emph{What is the best Sichuan restaurant neighbouring}. (detailed profile and corresponding queries are list in App.\ref{sec:app:reflect:personal-query}).
The personal query set is used as input for \Reflect{} without \Init{} or \Retrieve{}, we collect output of \Reflect{} ($\reflection$) for evaluation.
Following similar procedure in Sec.\ref{sec:experiments:personalized-response}, similarity between embedded $\reflection$ and embedded initial profile $\profile$ is computed to measure whether \Reflect{} is able to capture information from personal query.

Specially, we collect 11 users and generate 10 queries for each user, forming a 10-turn conversation for each user and leading 10-turn reflection. After reflection in each turn, we compute the cross similarity between current reflection and initial profile. 
Similar as in Sec.\ref{sec:experiments:personalized-response}, the we measure the Top1 accuracy of personalization by determining whether the top similarity reflection belong to the corresponding profile. The accuracy is regarded as the performance of \Reflect{} in each turn. As multi-turn conversation progress, a well-defined \Reflect{} is supposed to demonstrate increasing accuracy for personalization.

We further measure the similarity between initial profile and corresponding personal queries as upper bound: As personal queries are information abstraction from initial profile while reflection is another level abstraction from queries, it is expected that the similarity between reflection and initial profile should be lower than that between queries and initial profile.


Performance of \Reflect{} is dominated by its underlying LLM module. We use Baichuan2, GPT-3.5, GPT-4.0 for comparison. Detailed prompt, personal query and reflection is lised in App.\ref{sec:app:reflect}).

\begin{figure}[!htbp]
    \centering
    \includegraphics[width=\textwidth]{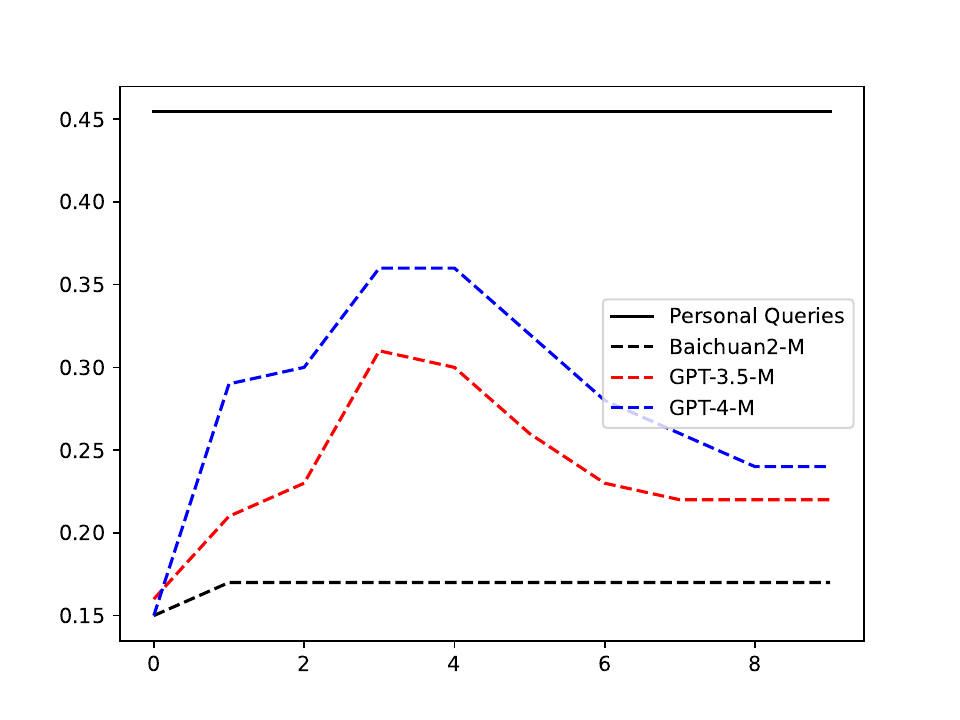}
    \caption{Personalization (Top 1) of reflection w.r.t initial profile as conversation progresses. Reflection is generated for each query and personalization is computed after each reflection. Personalization of query set and initial profile is plotted above as an upper line.}
    \label{fig:exp:reflect:tendency}
\end{figure}

As shown in Figure \ref{fig:exp:reflect:tendency}: 
Personalization between query set and initial profile is above $45\%$, showing that the personal query set captures and reflects the properties of user.
On average, GPT-series demonstrates large margin outperformance than Baichuan2: Baichuan2 shows an averaged accuracy around $17\%$ for 100-classification problem. 
However, GPT-3.5/4.0 achieves above $30\%$ in the same setting.
Besides, GPT-series do not show increasing personalization as conversation progresses.
It is assumed that users' diversity is not fully exposed in query set, leading to similar reflection.

%% file: content/appendix.tex
\subsection{\Init{}}\label{sec:app:init}
\subsubsection{Prompt}\label{sec:app:init:prompt}
\begin{mdframed}[backgroundcolor=gray!20, linecolor=black, linewidth=2pt, roundcorner=5pt]
As an experienced data analyst and synthesizer, given a user's behavior sequences: json \{\}, you need to summarize and inference the user's specific preferences. During summarizing you should obey following procedures: Firstly, please compile an overview of the user's daily consumption in this scenario in details, for example, where the user prefers to travel in their daily life, how often? so that one can fully understand the user's all-around habits without further access to their records. Next, identify and summarize the common characteristics of the consumed items, such as whether the movies feature the same actors, share similar themes or the attractions share similar style and so on. Finally, speculate on how this might reflect the user's life or state of mind. Please think step by step and show short inference summarization. Finally output the final results according to the following json format as: json \{\}.
\end{mdframed}
\subsubsection{Examples: User Initial Profile}\label{app:response:example1:profile}
We list an example of user profile, together with their raw data in Meituan:
\paragraph{Interactions}
\begin{mdframed}[backgroundcolor=gray!20, linecolor=blue, linewidth=2pt, roundcorner=5pt]
\{"Take Away Home": ["Ziyan Barilla Chicken-Snack \& Duck Neck-Beijing", "Gaoji Liangpi \& Skewers-Snack \& Dine-Fried Chicken \& Skewers-Tianjin", "AVOCADO TREE-Western Food-Salad-Beijing", "Xiaoxia Xiangfu Shanghai Cuisine-Beijing", "Parisian Desserts-Bread \& Desserts-Bakery \& Cake-Beijing", "Tingzhen Beef Noodle-Snack \& Rice Noodle-Beijing", "Tims Tianhao Coffee-Cafe-Beijing", "Eat Yixin Bird Nest-Food Supplements-Supplements-Beijing", "Eat Yifang Stewed Bird Nest-Food Supplements-Beijing", "Tims Sky Good Coffee-Cafe-Beijing", "Eat Yixin Stewed Bird Nest Fast Food - Rice Vermicelli - Beijing", "Tims Tianhao Café - Café - Beijing", "Food Yixian Bird's Nest - Food Tonic - Tonic - Beijing", "Hundred Years of Marinade - Snacks - Old Beijing Snacks - Beijing", "Jelly Duck Neck - Snacks - Marinade Duck Neck - Beijing", "Frog Lai Da - Hunan Cuisine - Beijing", "Burger King - Snacks - Western Fast Food/Hamburger - Beijing ", "FOODBOWL SUPER BOWL-Snacks - Fast Food/Light Meals-Beijing", "Blueglass-Yoghurt \& Fresh Milk-Beijing", "Guifeng Guizhou Rice Noodles-Other Regional Cuisine Guizhou Cuisine|Guizhou Cuisine-Beijing", "Xing Yue-Cantonese Cuisine-Barbecue-Roasted Meat-Beijing", "SUBWAY-Snacks-Western Fast Food/Sandwiches-Beijing", "Crossing the Bridge Edge Over Bridge Rice Noodle - Snacks - Fast Food - Beijing", "Big Duck Pear Roast Duck Restaurant - Beijing Cuisine - Roast Duck - Beijing", "Naixue's Tea - Milk Tea/Juice - Beijing", "In the Heavenly Mountains - Xinjiang Cuisine - Xinjiang Cuisine - Beijing", "Kudi Coffee - Coffee Shop - Beijing", "Kuaifu Frying Skewers - Snacks - Fried Chicken \& Frying Skewers - Beijing", "A-Shang-A-Shang Spicy Hotpot and Hot Spicy Hot Pot - Snacks - Hot \& Spicy - Beijing", "Long Xi Liao Pork - Snacks - Deli \& Smoked Sauce - Beijing", "Half Day Demon Grilled Fish - Szechuan Cuisine - Grilled Fish - Beijing", "AmazingGreen Low-Calorie - Light Meals - Western Cuisine - Light Salads - Beijing", "Costa Coffee - Café - Beijing", "Shano Light Meals - Western Cuisine - Light Salads - Beijing", "Big Round Bowl - Small Eat Fast Food - Fast Food Simple Food - Tianjin"], "Restaurant to Shop": ["Tube's Wings Bar-Barbecue-Wings-Beijing", "Tube's Wings Bar-Barbecue-Skewers-Beijing", "Xita Old Lady Mud Hearth Barbecue-Barbecue-Korean Barbecue-Beijing"], "Movie Shows": ["Movie Title: Saving the Suspect; Movie Genre: Suspense,Crime, Drama; Movie Starring: Zhang Xiaofei,Li Hongqi, Hui Yinghong,Wang Ziqi,Hong Junjia. : Movies", "Movie Title: The Hunger Games: Song of the Songbird and the Serpent; Genre: Action,Sci-Fi,Adventure; Movie Starring: Tom Bryce,Rachel Zeigler,Pete Dinlage,Jensen Schwartzman,Josh Andres Rivera. : Movies"], "Lifestyle Shopping": ["Xin Xing Lin Beef Jerky 45g: Dried Meat/ Beef Products/", "Shrimp Slider : Freshly Prepared Fresh Food/ Kanto Boiled/", "Bamboo Shoots Tips : Rootstalks/ Bamboo Shoots/", "Kasi Sugar Break Diary Original Yoghurt 130g : Yoghurt/Live Bacteria Yoghurt (Low Temperature)/", "Yellow Glutinous Corn Approx. 200g: Cooked Food/ Side Dietary/ Cooked Corn/", "Kasi Sugar Break Diary Original Yoghurt 130g : Yoghurt/Live Bacteria Yoghurt (Low Temperature)/", "Xin Xing Lin Beef Jerky 45g: Dried Meat/ Beef Products/", "Shuang Hui Crispy Sausage 140g: Convenient Instant Food/ Instant Sausage/", "Coca-Cola Heroes' Debut Soft Drink 330ml: Drinks/Carbonated Drinks/", "Sweet \& Spicy (Chinese Good Stew) : Instant Frozen Side Dish/Frozen Meatballs/", "Good Stew Hard-boiled Eggs 50g. Cooked Food / Side Dish / Cooked Eggs/", "Shrimp Slider : Freshly Prepared / Kanto Boiled/", "Charcoal Grilled Leg Steak 120g New : Freshly Prepared / Deep Fried Snacks/", "Yellow Glutinous Corn Approx. 200g: Cooked Food / Side Dish / Cooked Corn/", "White Jade Red Bean \& Job's Tears Soya Bean Milk Bottled 280ml: Beverages / Plants / Complex Protein Drinks/", "Xin Xing Lin Beef Jerky 45g: Jerky \& Dried Beef / Beef Products/", "Beef Tendon D14 : Dried Meat / Beef Products/", "Octopus Cake : Freshly Prepared Food / Kanto Cooked Food/", "Yida Fruit Dandelion Skin 200g : Dried Fruit Preserves / Hawthorn Products/", "Potted Skewered Meat (Chicken) 100g : Freshly Prepared Food / Deep Fried and Baked Snacks/", "Thick Cut Bean Curd 30g : Marinated and Spicy Food / Soya Bean Products/", "Jieshite Cleaning Wet Paper Tissues Ice Feeling 15 Pieces: Paper Wet Wet Wipes / Wet Paper Tissues /Wet Toilet Paper/", "Shiitake Mushroom Skewers (Chinese Stew) : Mushrooms / Shiitake Mushrooms/", "Sea River Cocoa Milk 220ml : Milk / Goat's Milk / Flavoured Milk / Prepared Milk /", "Tofu Skewers : Ready-to-eat / Kanto Boiled /", "Radish H New : Ready-to-eat / Kanto Boiled /", "Vegetarian Tripe (Chinese Stew) : Ready-to-eat / Kanto Boiled /", "Shredded Small Knots of Konjac: Ready-to-eat / Kanto Boiled /", "Chenguang Sealing Tape 1 roll: Office supplies / Services / Tape / ", "Fruit ripe Gardenia Gardenia Oolong Gardenia Oolong Tea 500ml: Beverage / Tea Drinks / "], "Leisure and Entertainment": ["Barbie Hall Animal Hospital-Pet Hospital-Beijing: Pets / Pet Hospital / ", "Dora \& Doll Dora \& Dolls-Game Room-Doll House-Beijing: Leisure and Entertainment/Game Room/Doll House ", "Recreation \& Play Card-Game Hall-VR-Shanghai: Leisure \& Entertainment/Game Hall/Arcade"], "Medical \& Beauty": ["Lillian/Lillian/Beauty \& Beauty", "Lillian/Lillian/Beauty \& Beauty/Skin Management", "Medical \& Health/Checkup Centre"], "Medical \& Health": [], "Travel \& Accommodation": ["Bubble Mart City Park-Theme Park-Theme Park-Beijing - Beijing - Beijing Municipality:"], "Travel \& Stay": ["Bubble Mart City Park-Theme Park-Beijing - Beijing Municipality:"], "Travel \& Stay": ["Travel \& Stay"], "Travel \& Stay": ["Bubble Mart City Park-Theme Park-Theme Park-Beijing - Beijing Municipality. Travel"]\}
\end{mdframed}
\paragraph{Profile}
\begin{mdframed}[backgroundcolor=gray!20, linecolor=blue, linewidth=2pt, roundcorner=5pt]
Lifestyle Product Preferences: The user's purchase records show that they have bought various types of lifestyle products, including food, beverages, office supplies, and personal care products. Within the food category, the user has purchased various types of meat products, such as beef jerky and crispy sausages, as well as a variety of fresh and cooked foods. Additionally, the user has bought various drinks, including yogurt, carbonated drinks, plant-based protein beverages, and tea drinks. In the personal care category, the user has bought facial cleansing wipes. In the office supplies category, the user has bought packing tape. These purchase records indicate that the user has a broad range of lifestyle product needs, covering various aspects of daily life. The user may be someone who likes to try new things, as their purchasing behavior shows an interest in a wide variety of product types.

Entertainment and Travel Preferences: The user's everyday leisure activities are mainly focused on arcades and pet hospitals, with spending records at the arcade including dollhouses and VR game cities, showing the user's love for gaming and electronic entertainment. At the same time, the user also \ul{pays attention to their pet's health}, regularly visiting pet hospitals for check-ups. In terms of travel, the user has chosen \ul{theme parks} as their travel destination, likely preferring creative and fun travel experiences. Overall, the user's entertainment and travel preferences show characteristics of being lively, fun-loving, and caring for pets.

User Health Status: The user's behavior sequence is missing, and their preferences cannot be inferred.

Movie and Performance Preferences: The user has watched two movies, namely ``Saving the Suspect'' and ``The Hunger Games: The Ballad of Songbirds and Snakes.'' ``Saving the Suspect'' is a mystery, crime, and drama genre movie, starring Zhang Xiaofei, Li Hongqi, Hui Yinghong, Wang Ziyi, and Hong Junjia. The Hunger Games: The Ballad of Songbirds and Snakes'' is an action, science fiction, and adventure genre movie, starring Tom Blyth, Rachel Zegler, Peter Dinklage, Jason Schwartzman, and Josh Andrés Rivera. From these two movies, the user may prefer films with rich plots and diverse genres, and they do not have a particular preference for any one actor.

Culinary Preferences: The user's culinary preferences are extensive, covering multiple categories including snack fast food, Western cuisine, Hunan cuisine, bread desserts, food tonics, Beijing cuisine, milk tea juice, Xinjiang cuisine, Sichuan cuisine, and more. In the snack fast food category, the user prefers braised flavors like duck neck, fried chicken skewers, rice noodles, old Beijing snacks, \ul{Western fast food/burgers}, fast casual meals, spicy hot pot, smoked and marinated foods, etc., with the most frequented brands being Ziyanyibaiweiji, Gaoji Liangpi Fried Skewers, Tingzhen Beef Noodles, Bainian Luzhu, Juewei Duck Neck, Burger King, FOODBOWL Super Bowl, Guoqiaoyuan·Cross-bridge Rice Noodles, Kuafu Fried Skewers, Ashang Ashang Spicy Hot Pot, Longxi Lao Meat, and Dayuan Bowl, among others. In the Western cuisine category, the user prefers light food salads, with frequently visited brands including Avocado Tree Restaurant, AmazingGreen Low Carbism·Light Food, and Shaye Light Food. In the beverage category, the user prefers coffee and milk tea juice, with frequently visited brands including Tims Tianhao Coffee, Nayuki's Tea, Kudi Coffee, and COSTA COFFEE. The user's food choices show that they like to try various flavors of food, are not limited to a single taste, and also like healthy light food and beverages.

Medical Beauty/Beauty Preferences: The user's consumption sequence shows that their spending on medical beauty/beauty projects is mainly focused on 'beauty and body care' and 'skin management.' This may reflect the user's high demand and attention to maintaining a good appearance and skin condition. The user may often engage in these projects in daily life to maintain their image and figure. Additionally, the user also has records of spending at 'health check-up centers,' which may indicate that they also pay attention to their health condition. Overall, the user may have ongoing needs for beauty and body care and skin management, and they also pay attention to their health status.

\end{mdframed}

\subsection{\Response{}}\label{sec:app:response}
\subsubsection{Prompt}\label{sec:app:response:prompt}
To keep \method{} simple as a framework, we use the following response prompt:
\begin{mdframed}[backgroundcolor=gray!20, linecolor=black, linewidth=2pt, roundcorner=5pt]
\emph{You are a chatbot, based on the user's historical conversation, the current user's input, and according to the user's profile $\langle\lbrace \text{profile} \rbrace \rangle$, you engage in conversation with the user.}
\end{mdframed}
\subsubsection{Subjective Query List}\label{sec:app:sub-query}
LLM is tending to provide objective result for impersonal query, such as ``What is the highest mountain on earth?'', ``Tell me something about World War II.''
To evaluate personalization ability of \method{}, we collect some subjective queries to induce personal response. 
The queries are corresponding to user initial profile (Sec.\ref{sec:method:init}), which is categorized into 
``Lifestyle Product Preferences'', 
``Entertainment and Travel Preferences``, 
``User Health Status'', 
``Movie and Performance Preferences'',
``Culinary Preferences'',
``Medical Beauty/Beauty Preferences'':
\begin{mdframed}[backgroundcolor=gray!20, linecolor=blue, linewidth=2pt, roundcorner=5pt]
    \begin{itemize}
        \item What kind of movies do I like to watch?
        \item What do I like to eat?
        \item What do I like to do in my spare time?
        \item How has my health been recently?
        \item What's good to eat for lunch?
    \end{itemize}
\end{mdframed}
\subsubsection{Examples: Initial Profile \& Query \& Response}\label{app:response:example1}
Some examples are listed with initial profile and corresponding response using Baichuan2, GPT-3.5 and GPT-4:

\paragraph{Query}
What kind of movies do I like to watch?
\paragraph{Response}
\subparagraph{Baichuan2}
\begin{mdframed}[backgroundcolor=gray!20, linecolor=green, linewidth=2pt, roundcorner=5pt]
\ul{I do not have a favorite movie genre}. I was designed for natural communication with humans, to answer questions, assist in creation, and help the public easily and inclusively access world knowledge and professional services. If you have any questions, feel free to ask me at any time.
\end{mdframed}
Baichuan2 mistakes the query as asking chatbot itself and provides a response that it does not have movie preference. Though user profile is provided, it fails to involve the information to provide personal response.

\subparagraph{GPT-3.5}
\begin{mdframed}[backgroundcolor=gray!20, linecolor=green, linewidth=2pt, roundcorner=5pt]
\ul{According to your purchase history}, you have watched "Saving the Suspect" and "The Hunger Games: The Ballad of Songbirds and Snakes," which are mystery crime and action sci-fi adventure movies, respectively. It appears that \ul{you are interested in movies with rich plots and a variety of genres, without a particular preference for any one type or actor}. If you wish to try other types of movies, you can tell me the movie themes you are interested in or the movie genres you want to explore, and I can recommend some movies for you.
\end{mdframed}
GPT-3.5 is able to recall user profile and generate reasonable recommendation results with reasons.

\subparagraph{GPT-4}
\begin{mdframed}[backgroundcolor=gray!20, linecolor=green, linewidth=2pt, roundcorner=5pt]
\ul{Based on your past movie-watching records}, you have watched a diverse range of movie genres, including mystery, crime, drama, action, science fiction, and adventure. Specifically, you have watched "Saving the Suspect" and "The Hunger Games: The Ballad of Songbirds and Snakes." \ul{This indicates that you may prefer movies with rich plots and a variety of genres}.
\end{mdframed}
Similar as GPT-3.5, GPT-4 is able to capture information of user preference and give recommendation results. However, it does not show significant difference.

\subsubsection{Cross Personalization: Response w.r.t. Initial Profile}
During each query and response, we can generate a 2-dimension cross personalization heatmap between initial profile and response, elements in $(i,j)$ represents the $\PersonalizedMetric$ for initial profile of user $i$ and response of user $j$. 
A well-conditioned dialog system should generate a diagonal heatmap: the diagonal elements possess the highest personalization value while the rest are low.
We average the cross personalization heatmap for 5-turn conversations for the 11 users, and visualize the heatmap for a clear comparison:
\begin{figure}[!htbp]
  \centering
  \begin{subfigure}[b]{0.3\textwidth}
    \includegraphics[width=\textwidth]{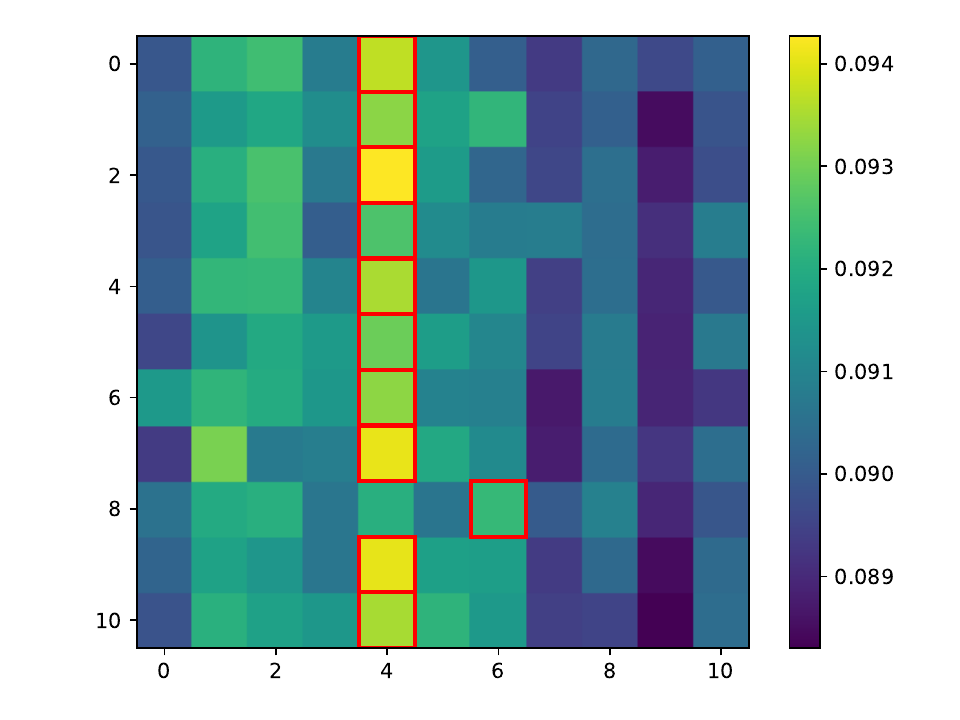}
    \caption{Baichuan2}
    \label{fig:app:response:baichuan2}
  \end{subfigure}
  \hfill 
  \begin{subfigure}[b]{0.3\textwidth}
    \includegraphics[width=\textwidth]{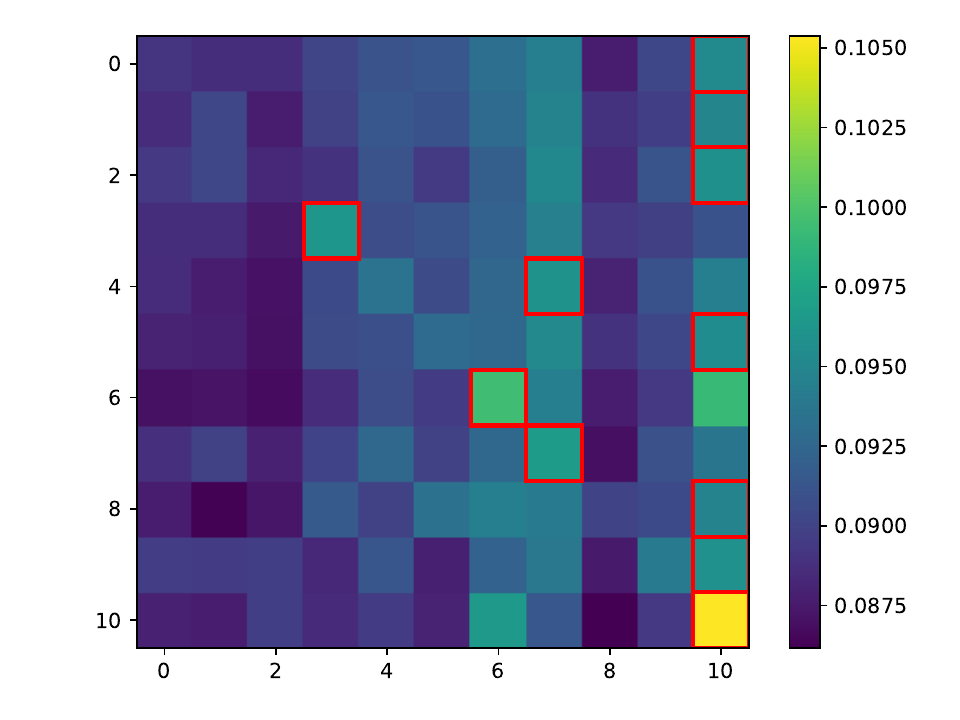}
    \caption{GPT-3.5}
    \label{fig:app:response:gpt_3_5}
  \end{subfigure}
  \hfill 
  \begin{subfigure}[b]{0.3\textwidth}
    \includegraphics[width=\textwidth]{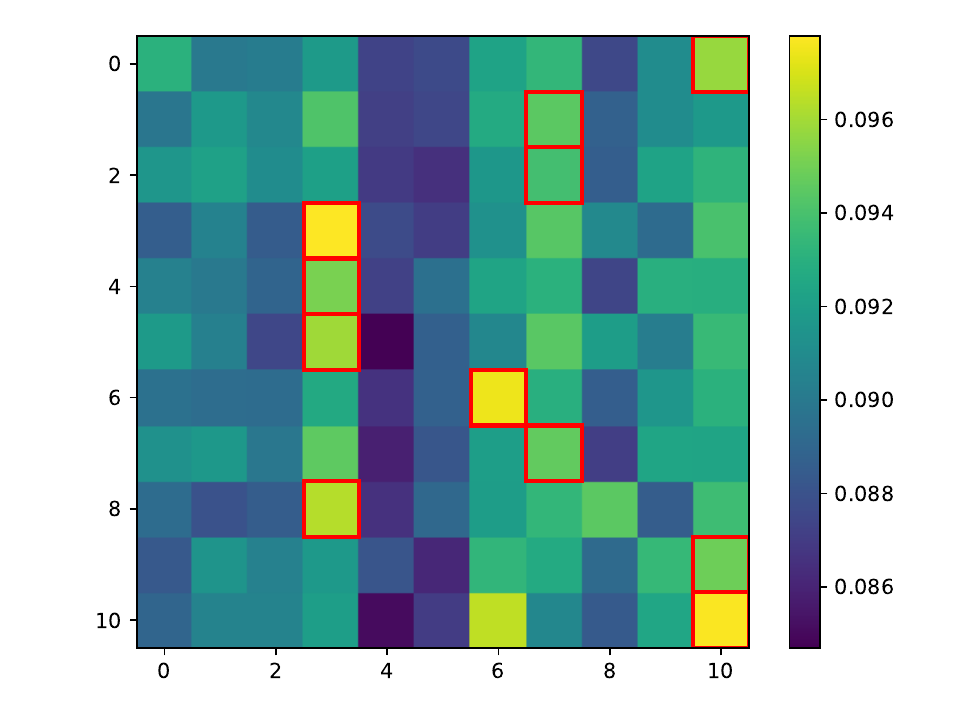}
    \caption{GPT-4}
    \label{fig:app:response:gpt_4}
  \end{subfigure}
  \caption{Cross personalization between initial profile and response. Row $i$ represents personalization between initial profile of user $i$ and all responses. Softmax is applied in each row, and the maximum element is bounded in red. A well-designed dialog system should produce large value for element in ($i,i$), leading to a heatmap with high value in diagonal elements. It means that user $i$'s response is highly related to corresponding initial profile. Personalization is provided enough to distinguish users.}
  \label{fig:app:response:cross-personalization}
\end{figure}

Fig.\ref{fig:app:response:cross-personalization} shows average cross-personalization between 11 users in 3 experiments (Baichuan2, GPT-3.5, GPT-4):

\paragraph{Baichuan 2}
In Fig.\ref{fig:app:response:baichuan2}, most initial profiles resemble $4$-th response. A example (Query: \emph{What kind of movies do I like to watch?}) of response from $4$-th user is listed:
\begin{mdframed}[backgroundcolor=gray!20, linecolor=green, linewidth=2pt, roundcorner=5pt]
\begin{itemize}
    \item I do not have any food preferences. I am designed to engage in natural communication with humans, answer questions, assist in creation, and help the general public easily and inclusively access world knowledge and professional services. If you have any questions, feel free to ask me at any time.
    \item I do not have a favorite type of movie. I am designed to engage in natural communication with humans, answer questions, assist in creation, and help the general public easily and inclusively access world knowledge and professional services. If you have any questions, you can ask me at any time.
\end{itemize}
\end{mdframed}
It is obvious that the response from $4$-th simply emphasizes that it is a chatbot and can not provide any personal preference. However, the query aims at responding user's preference based on initial profile. Baichuan2 fails to solve the task and generate a totally neutral response. However, such utterance generate a high personalization value.

\begin{figure}[!htbp]
    \centering
    \includegraphics[width=\textwidth]{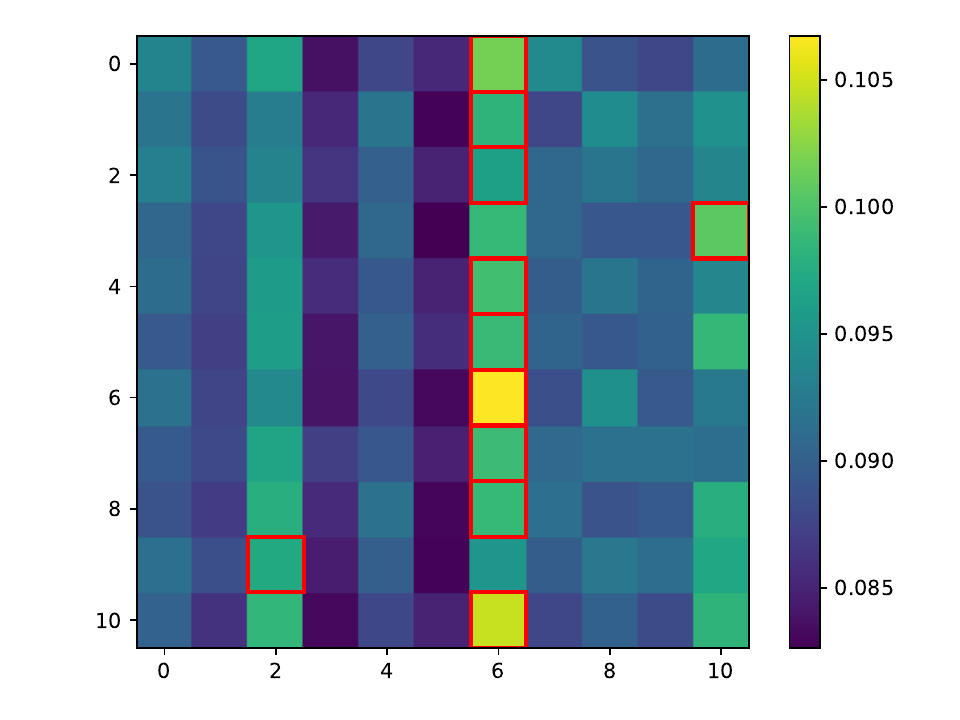}
    \caption{Cross personalization under query \emph{What kind of movies do I like to watch?}. Most profiles' personalization collapse to $6$-th's response.}
    \label{fig:app:response:cross-personalization:query_0}
\end{figure}

\paragraph{GPT-3.5}
GPT-3.5 generates partial diagonal pattern as shown in Fig.\ref{fig:app:response:gpt_3_5}. 
However, it is still noticed that most initial profiles agree that response for $11$-th user is the most match. 
\subparagraph{A Bad Case}\label{sec:app:response:cross-persoanlization:bad-case}
We list the top $3$ response (together with personalization score and corresponding initial profile) for $4$-th initial profile (App.\ref{app:response:example1:profile}) under query \emph{What kind of movies do I like to watch?}:
\begin{mdframed}[backgroundcolor=gray!20, linecolor=green, linewidth=2pt, roundcorner=5pt]
\begin{itemize}
    \item [$6$] [$0.09945668848188244$] Based on your historical purchase records and your recent viewing of the movie ``Dad's School'', it can be inferred that you may like movies with rich storylines and family elements. At the same time, you also seem to have a certain preference for actors such as Huang Bo, Shan Yuhao, Yan Ni, Zhang Junning, and Zhang Zixian. Therefore, the characteristics of movies you may like include rich plots, family elements, and starring actors like Huang Bo, Shan Yuhao, Yan Ni, Zhang Junning, and Zhang Zixian.
    \item [$2$] [$0.09591810694688538$] Based on your historical purchase records, it appears that you have a preference for romance and comedy movies. These two genres have been watched by you multiple times, which may indicate a higher preference for these types of films. Of course, this does not mean that you do not like other genres of movies. If you have other specific preferences or would like to know recommendations for other types of movies, please let me know, and I will provide more detailed suggestions.
    \item [$10$] [$0.09376776688012621$] Based on your purchase history and viewing preferences, you may favor drama and comedy movies. Among the movies you've watched, 'Second Hand Masterpiece' and 'Not That Fired Up' are categorized as dramedy. Additionally, you might have a particular liking for actor Yu Hewei's performances, as he stars in both 'Second Hand Masterpiece' and 'Solid as a Rock'. I hope this information is helpful to you! If you have any other questions, feel free to ask at any time.
\end{itemize}
\end{mdframed}
Profile in App.\ref{app:response:example1:profile} clearly mentions about movie "Saving the Suspect"/"The Hunger Games: The Ballad of Songbirds and Snakes," and movie types about mystery crime and action sci-fi adventure. However, the response with high score does not mention about the related content. It shows a bad case of $\PersonalizedMetric$. Fig.\ref{fig:app:response:cross-personalization:query_0} shows the cross-personalization heatmap for this query

\subparagraph{A Good Case}
Though users are undistinguishable in previous case. Average cross personalization is able to distinguish $4$-th user as shown in Fig.\ref{fig:app:response:gpt_3_5}. We show another example on query \emph{What do I like to do in my spare time?}:
\begin{figure}[!htbp]
    \centering
    \includegraphics[width=\textwidth]{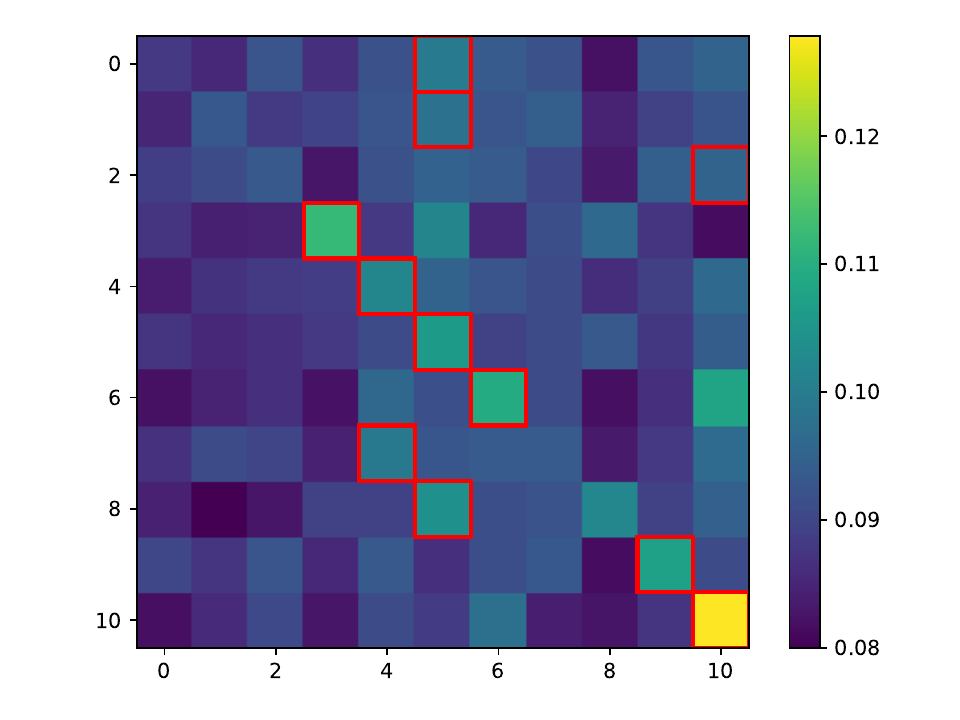}
    \caption{Cross personalization under query \emph{What do I like to do in my spare time?}. A partial diagonal pattern is obvious, showing that the dialog system is able to distinguish users under the query.}
    \label{fig:app:response:cross-personalization:good-case}
\end{figure}
Fig.\ref{fig:app:response:cross-personalization:good-case} visualizes cross personalization on the query. Top $3$ response for $4$-th initial profile is listed:
\begin{mdframed}[backgroundcolor=gray!20, linecolor=green, linewidth=2pt, roundcorner=5pt]
\begin{itemize}
    \item [$4$] [$0.10194599645809517$] Based on your purchase records and consumer preferences, the following conclusions can be drawn: you like to try various types of daily necessities and food, and you enjoy purchasing food, beverages, office supplies, and personal care products. In terms of food, you have bought various meat products, fresh-made meals, and ready-to-eat foods. For entertainment, you like to visit arcades and pet hospitals. You also choose theme parks as travel destinations for creative and fun experiences. Additionally, you have watched different types of movies, indicating an interest in films with rich plots and diverse genres. You also enjoy trying a variety of cuisines, not limiting yourself to one particular taste, and you have a certain preference for healthy light meals and beverages. You have also spent on medical beauty/personal care services, paying attention to your appearance, physique, and skin condition. In summary, you have a wide range of activities, like to try new things, and are concerned with maintaining your appearance and health.
    \item [$10$] [$0.0963293581109227$] Based on your purchase records and profile analysis, you may be someone who enjoys cooking at home, studying or working, and pays attention to personal emotional regulation. You have purchased various types of daily necessities, including food, stationery, cooking ingredients, and personal care products. The food items you bought mainly consist of various vegetables, soy products, eggs, and dried goods, which may indicate that you like to cook at home. You also bought some stationery items, such as highlighters and notebooks, which may suggest that you have study or work needs. Additionally, you purchased some cooking ingredients, like flour and sugar, which may indicate that you enjoy baking. Lastly, you bought personal care items like flowers, which may suggest that you focus on regulating your emotions. Overall, you may enjoy activities such as cooking at home, studying or working, and paying attention to personal emotional regulation.
    \item [$5$] [$0.09520748525028173$] Based on your purchase history and profile analysis, you may be a modern individual with a fast-paced lifestyle, who values personal appearance and comfort. You likely prefer a convenient and efficient way of living, paying attention to life's details and personal comfort. Regarding daily necessities, you have purchased fruits, oral care products, men's grooming items, drinking water, spicy marinated foods, ready-to-eat meals, beauty tools, and fashion accessories, indicating a wide range of needs for daily goods. You may favor pre-cut fruits, care about personal image and hygiene, and need quick meal solutions.

    In terms of entertainment and travel, you enjoy singing at KTV, particularly at wholesale-style KTVs, and you often engage in this activity in Shijiazhuang city. When it comes to travel, you prefer to stay in accommodations, with a particular preference for four-star/luxury hotels, and you have stayed multiple times in Shanghai and Shijiazhuang cities. Regarding health, you may have some issues with your eyes, oral care, and skin, as well as some needs related to family medicine and sexual health. In summary, you enjoy a lifestyle that is convenient, quality-oriented, and focused on personal image.
\end{itemize}
\end{mdframed}

\paragraph{GPT-4}
Diagonal pattern is becoming obvious in Fig.\ref{fig:app:response:gpt_4}. Similar cases happen in GPT-4 as in GPT-3.5.

\subsection{\Retrieve{}}\label{sec:app:retrieve}
\subsubsection{Prompt for LLM based Retrieval}
\begin{mdframed}[backgroundcolor=gray!20, linecolor=black, linewidth=2pt, roundcorner=5pt]
\emph{Based on the user's input $\langle \lbrace \text{input} \rbrace \rangle$, extract relevant content from the user's profile $\langle \lbrace \text{profile} \rbrace \rangle$. Identify the topic of user input, such as dietary preferences, interests, etc. Then extract content related to the judged topic from the profile, avoiding irrelevant details. Summarize concisely, not exceeding 50 words. Your output should only include the extracted content from the profile, without the reasoning process.}
\end{mdframed}

\subsubsection{Examples: Retrieval Profile \& Response}
We use the same user profile and query in App.\ref{app:response:example1}, and show the retrieval profile using embedding based and LLM based method, together with their corresponding response:
\paragraph{Embedding based Retrieval}
\begin{mdframed}[backgroundcolor=gray!20, linecolor=blue, linewidth=2pt, roundcorner=5pt]
Movie and Performance Preferences: The user mainly watches movies in the genres of drama, comedy, crime, and action, with a larger proportion of drama and comedy films. Among the actors, Lei Jiayin, Yu Hewei, and Zhang Songwen appear frequently in the movies watched by the user. Based on the types of movies and actors preferred, the user likely enjoys films with rich plots and depth, as well as light-hearted comedies. In terms of actor preference, the user may favor the acting styles of Lei Jiayin, Yu Hewei, and Zhang Songwen.
\end{mdframed}
Embedding based retrieval is able to retrieve movie-related profile. This profile generates the following response:
\begin{mdframed}[backgroundcolor=gray!20, linecolor=green, linewidth=2pt, roundcorner=5pt]
Based on the two movies you've previously watched, ``Saving the Suspect'' and ``The Hunger Games: The Ballad of Songbirds and Snakes'', it seems you are interested in movies with rich plots and a variety of genres. You do not have a particular preference for any actor. Therefore, I will recommend some movies with rich plots and diverse genres that you might enjoy.

``Inception'': This is an exciting sci-fi movie that revolves around the complex relationship between dreams and reality, filled with mystery and thrill. It stars Leonardo DiCaprio, Joseph Gordon-Levitt, Ellen Page, and others.

``Source Code'': This movie combines sci-fi, mystery, and action, where the protagonist tries to identify a bomber in a never-ending 8-minute loop. It stars Jake Gyllenhaal, Michelle Monaghan, Jeffrey Wright, and others.

``The Bourne Identity'': This thrilling action movie tells the story of an amnesiac secret agent who uncovers a massive conspiracy while trying to discover his own identity. It stars Matt Damon, Franka Potente, Chris Cooper, and others.

``Se7en'': This is a crime mystery thriller that depicts the story of two detectives tracking a serial killer. It stars Morgan Freeman, Brad Pitt, Kevin Spacey, and others.

I hope these recommendations suit your taste, and if you need more, please let me know.
\end{mdframed}

\paragraph{LLM based Retrieval}
\begin{mdframed}[backgroundcolor=gray!20, linecolor=blue, linewidth=2pt, roundcorner=5pt]
Based on the user's purchase records, entertainment and travel preferences, and culinary tastes, the following can be concluded:

The user has bought various household items, likes to try new things, and is interested in different types of products.
The user enjoys arcades and pet hospitals, loves gaming and electronic entertainment, and cares about pet health.
The user chooses theme parks as travel destinations, preferring fun and creative travel experiences.
The user watches a variety of movie genres and likes movies with rich plots.
The user has a broad range of food preferences, including snacks, fast food, Western cuisine, and milk tea and fruit juices, and enjoys trying different flavors.
The user's spending on medical beauty/cosmetic services is mainly on beauty treatments and skin care, indicating a high demand for appearance and skin condition.
\end{mdframed}
LLM based retrieval generates more than movie-related information. Basically it summarizes the full profiles, which generates the following response:

\begin{mdframed}[backgroundcolor=gray!20, linecolor=green, linewidth=2pt, roundcorner=5pt]
Based on your purchase history and entertainment preferences, you enjoy watching movies with rich plots. This means you tend to prefer films with in-depth storytelling and emotional development. You may be interested in various types of movies and like to try new cinematic experiences. Therefore, you could try different genres of movies, such as \ul{romance, action, mystery, science fiction}, etc., to satisfy your diverse interest in films.
\end{mdframed}
Response from LLM based retrieval does not include details and recommendation as that of embedding based retrieval.

\begin{figure}[!htbp]
  \centering
  \begin{subfigure}[b]{0.45\textwidth}
    \includegraphics[width=\textwidth]{Figures/Experiments/Cross_Pers_Response_GPT_3_5.pdf}
    \caption{Full Profile}
    \label{fig:app:retrieve:full-profile}
  \end{subfigure}
  \begin{subfigure}[b]{0.45\textwidth}
    \includegraphics[width=\textwidth]{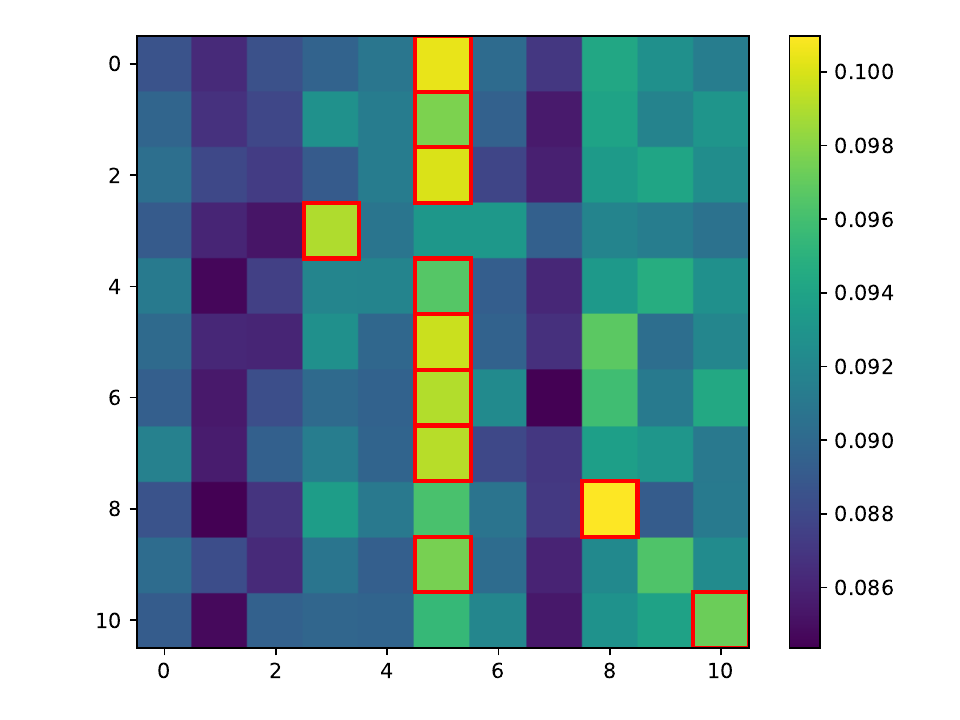}
    \caption{Embedding based}
    \label{fig:app:retrieve:embedding-based}
  \end{subfigure}
  \begin{subfigure}[b]{0.45\textwidth}
    \includegraphics[width=\textwidth]{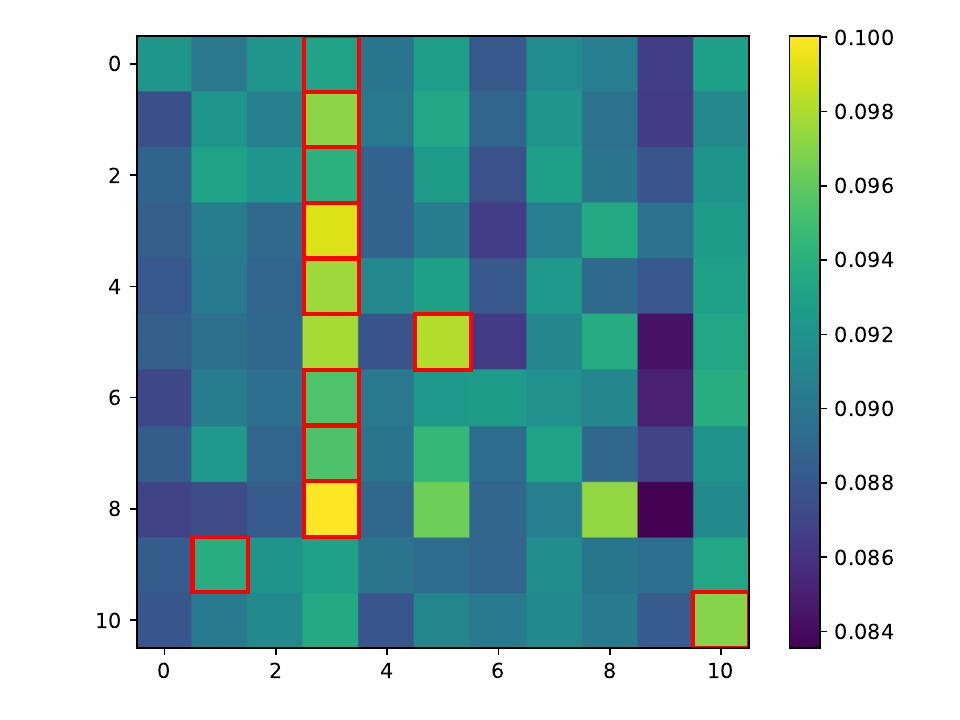}
    \caption{LLM based}
    \label{fig:app:retrieve:llm-based}
  \end{subfigure}
  \begin{subfigure}[b]{0.45\textwidth}
    \includegraphics[width=\textwidth]{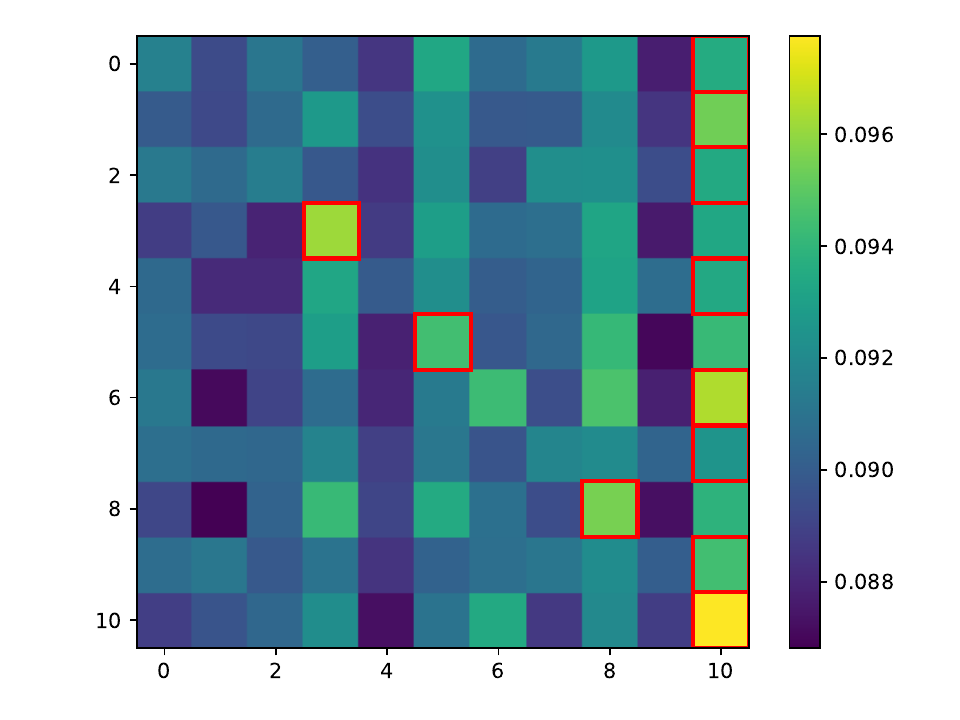}
    \caption{Embedding+LLM based}
    \label{fig:app:retrieve:embedding-llm-based}
  \end{subfigure}
  \caption{Cross personalization between initial profile and response under different retrieval methods.}
  \label{fig:app:retrieve:cross-personalization}
\end{figure}
\subsubsection{Cross Personalization: Response w.r.t Initial Profile}
\Retrieve{} aims at improving personalization by retrieving proper profile. In Fig.\ref{fig:app:retrieve:cross-personalization}, we visualize the average cross personalization between initial profile and response under different retrieval methods.

\subsection{\Reflect{}}\label{sec:app:reflect}
\subsubsection{Personal Queries Set}\label{sec:app:reflect:personal-query}
\begin{mdframed}[backgroundcolor=gray!20, linecolor=black, linewidth=2pt, roundcorner=5pt]
Based on the user profile, generate $\lbrace \text{num} \rbrace$ user queries. Overall, these queries need to reflect the content in the user profile, such as:
1. They should reflect user behavior; for example, if the profile mentions that the user frequently visits Western restaurants, then a query could be: ``Recommend some steak restaurants for me.''
2. They should reflect the user's personality; for instance, if the profile mentions that the user is cheerful, then the query should be expressed in an optimistic manner as much as possible.
3. They should not directly expose the user's information; they must express the profile content implicitly and subtly, like not directly saying ``I am a cheerful person,'' ``I've been going to Western restaurants frequently lately.''
User profile: $\lbrace \text{profile} \rbrace$.
\end{mdframed}
For example, LLM generates personal query for user in App.\ref{app:response:example1} as:
\begin{mdframed}[backgroundcolor=gray!20, linecolor=green, linewidth=2pt, roundcorner=5pt]
\begin{itemize}
    \item Recommend some fresh food and beverages for me.
    \item Is there a recommended pet hospital?
    \item I want to go to an interesting theme park.
    \item Are there any recommended mystery movies?
    \item Recommend some gourmet snacks.
    \item Is there a recommended Western restaurant?
    \item I want to try some healthy light food.
    \item Are there any recommended medical beauty treatments?
    \item I want to learn about some beauty and body care methods.
    \item Is there a recommended health check-up center?
\end{itemize}
\end{mdframed}
As we may notice in App.\ref{app:response:example1:profile}, profile mentions about user's travel to theme park, which brings a query on theme park. Besides, Western food is mentioned in profile, leading to a Western restaurant recommendation in the query.

\subsubsection{Prompts}\label{sec:app:reflect:prompt}
We use the following prompts for reflection generation:
\begin{mdframed}[backgroundcolor=gray!20, linecolor=black, linewidth=2pt, roundcorner=5pt]
You can infer information about a person from details, such as recent events, personality, dietary preferences. Given a user input, make an inference based on this information and provide the inferred information about this user. The inference process does not need to be output; reply as briefly as possible.
\end{mdframed}

\subsubsection{Examples}
We use the same user in App.\ref{app:response:example1} and corresponding query sets mentioned aboved and generate the following reflections:
\paragraph{Baichuan2}
\begin{mdframed}[backgroundcolor=gray!20, linecolor=green, linewidth=2pt, roundcorner=5pt]
Of course! Here are some fresh and healthy food and drink suggestions:

1. Food:
- Fruits and vegetables: Choose seasonal fruits and vegetables, such as strawberries, blueberries, cherries, tomatoes, cucumbers, etc. These foods are rich in vitamins and minerals, which help maintain health.
- Nuts and seeds: Such as almonds, walnuts, flaxseeds, and chia seeds. They are rich in healthy fats, proteins, and fiber.
- Lean meats: Such as chicken breast, turkey, fish, etc. They are rich in high-quality protein and other nutrients.
- Whole grains: Such as oats, brown rice, whole wheat bread, etc. They are rich in fiber, which helps with digestion and stabilizing blood sugar levels.
- Tofu and soy milk: They are good sources of plant-based protein and are also rich in calcium and vitamins.

2. Drinks:
- Water: It is very important to maintain adequate water intake. Try to choose pure water and avoid drinks with sugar or artificial ingredients.
- Green tea: Green tea contains antioxidants, which can help increase metabolism and reduce the risk of heart disease.
- Fruit juice: Try to choose 100\% pure natural fruit juice and avoid sugary fruit juice drinks. Juice can provide additional vitamin C and other nutrients.
- Milk and dairy products: Such as low-fat milk, yogurt, etc. They are good sources of calcium and protein, which help with bone health.
- Vegetable and fruit juices: Such as carrot juice, spinach juice, etc. They are rich in vitamins, minerals, and antioxidants.

Please note that everyone's nutritional needs may vary depending on factors such as age, gender, weight, lifestyle, etc. When adjusting your diet, be sure to consider personal needs and consult the opinion of a professional nutritionist.
\end{mdframed}
Instead of reflection, Baichuan2 basically performs conversation by responding to the query. Besides, its generation is too long as reflection and does not obey requirements in prompt.
Baichuan2 in unable to perform customized task without speical finetuning.

\paragraph{GPT-3.5}
\begin{mdframed}[backgroundcolor=gray!20, linecolor=green, linewidth=2pt, roundcorner=5pt]
Based on your interest in fresh foods and beverages, it can be inferred that you have high standards for health and taste. You may focus on consuming fresh fruits, vegetables, and organic foods, and enjoy trying novel juices, teas, or coffees. You likely have a high concern for both nutrition and taste.

The user may have pets and need to find a trustworthy veterinary hospital, possibly caring about the health and welfare of their pets.

This user may enjoy adventure and excitement, like trying new things, and could be an adventure enthusiast.

The user may be interested in mystery movies and likely enjoys deduction and puzzle-solving.

Based on the user's needs, some gourmet snacks can be recommended. It is speculated that the user has a certain interest in gourmet food and may like to try foods with different flavors.

The user may be interested in Western cuisine and willing to try new dining experiences.

This user may be paying attention to their health status, possibly wanting to control weight or improve dietary habits. They may have some understanding of healthy eating and are willing to try new foods or dishes.

Based on the user's inquiry about medical beauty recommendations, it can be deduced that the user may care a lot about their appearance and image, and has a high interest in personal image improvement and beauty needs.

This user may be conscious of their personal image and appearance, actively engaging in beauty and health care. They may be looking for ways to improve their appearance and physical condition.

This user may be concerned about their health and could also be preparing for a medical check-up.
\end{mdframed}
GPT-3.5 basically follows instruction in prompt and generate reflection for the user.

\paragraph{GPT-4}
\begin{mdframed}[backgroundcolor=gray!20, linecolor=green, linewidth=2pt, roundcorner=5pt]
You may be focused on a healthy diet, enjoying fresh food and beverages.

The user may be a pet owner looking for pet medical services.

This user may be someone who seeks fun and adventure, enjoying outdoor activities and interactive entertainment facilities.

This user likes to watch mystery movies.

The user may be a food enthusiast with a strong interest in trying new foods and snacks."

This user may like or want to try Western cuisine.

This user may be focusing on a healthy diet, possibly dieting or trying to lose weight.

This user may have a significant interest in beauty and maintenance, and also cares about their physical appearance. They are open to and actively seeking medical beauty solutions.

The user may be very concerned about maintaining a good appearance and physique, or may work in a profession related to beauty and body care.

This user may be concerned about their health or have a habit of getting regular health check-ups.
\end{mdframed}
GPT-4.0 generates more precise and tidy reflection.

\subsubsection{Cross Personalization: Reflection w.r.t Initial Profile}
\begin{figure}[!htbp]
  \centering
  \begin{subfigure}[b]{0.45\textwidth}
    \includegraphics[width=\textwidth]{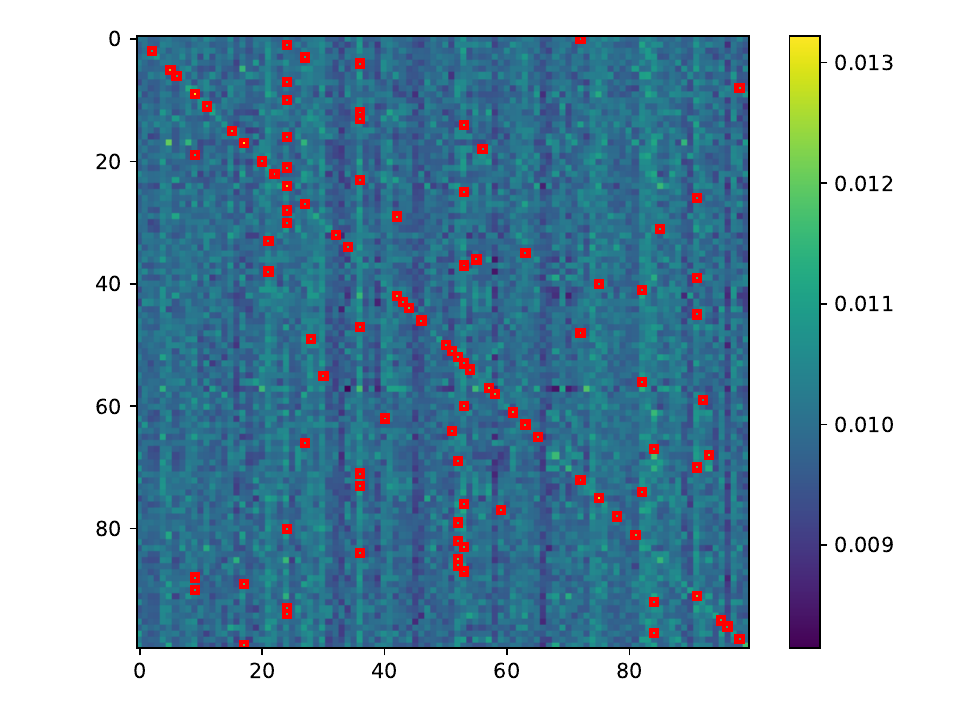}
    \caption{Query set}
    \label{fig:app:reflect:query-set}
  \end{subfigure}
  \begin{subfigure}[b]{0.45\textwidth}
    \includegraphics[width=\textwidth]{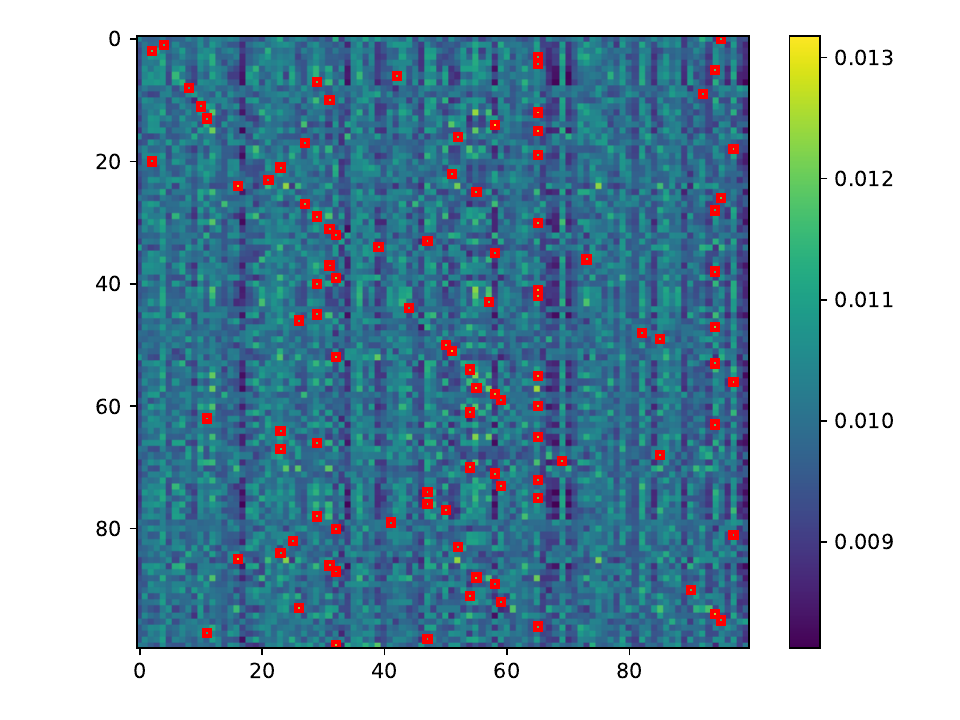}
    \caption{Baichuan2}
    \label{fig:app:reflect:baichuan2}
  \end{subfigure}
  \begin{subfigure}[b]{0.45\textwidth}
    \includegraphics[width=\textwidth]{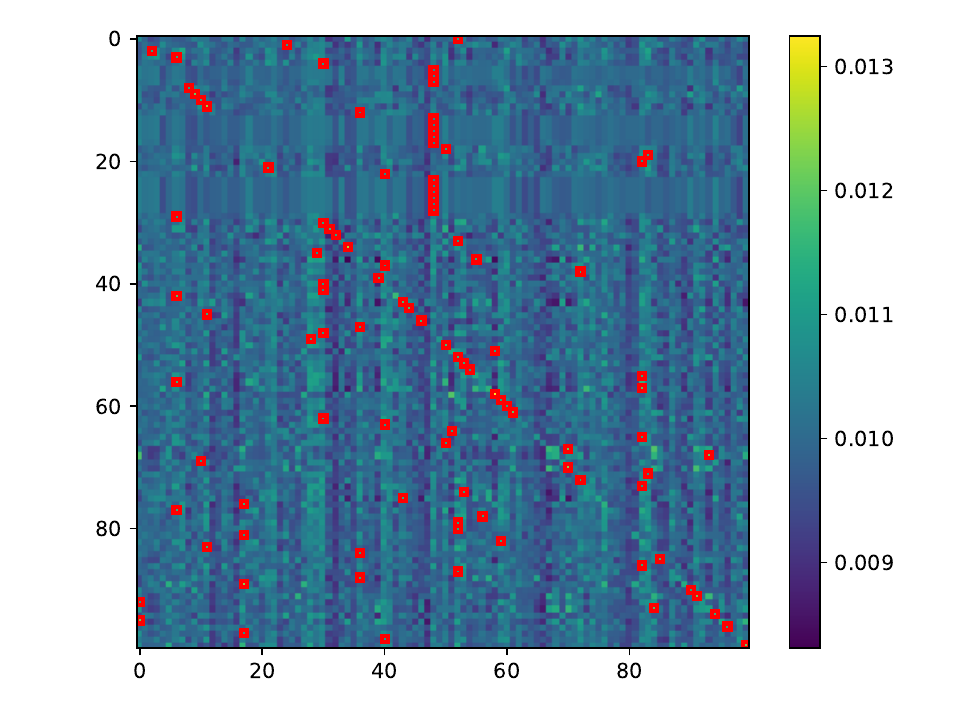}
    \caption{GPT-3.5}
    \label{fig:app:reflect:gpt-3-5}
  \end{subfigure}
  \begin{subfigure}[b]{0.45\textwidth}
    \includegraphics[width=\textwidth]{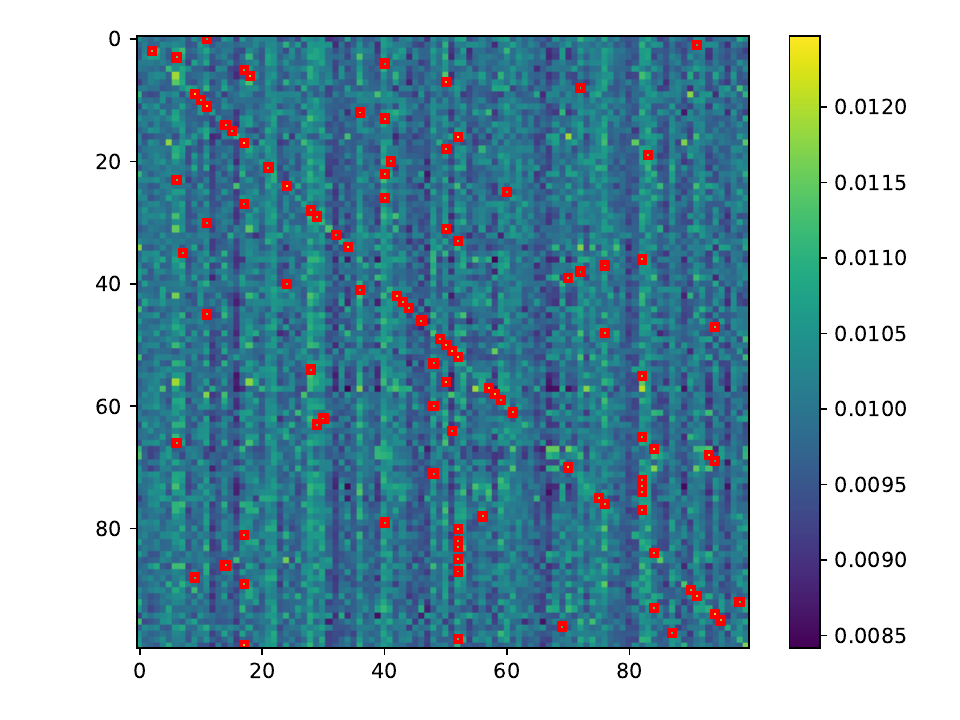}
    \caption{GPT-4}
    \label{fig:app:reflect:gpt-4}
  \end{subfigure}
  \caption{Cross personalization between initial profile and reflection under different reflection methods.}
  \label{fig:app:reflect:cross-personalization}
\end{figure}

\Reflect{} aims at generate personal reflection according to query sets. We visualize the average cross personalization between reflection and initial profile under different reflection foundation models in Fig.\ref{fig:app:reflect:cross-personalization}. Besides, a cross personalization on query set w.r.t initial profile is listed for comparison.

Fig.\ref{fig:app:reflect:query-set} shows personalization between query set and initial profile with a clear diagonal pattern. As shown in Fig.\ref{fig:app:reflect:baichuan2}, Baichuan's performance on reflection is weak, compared to that of GPT-3.5/4 in Fig.\ref{fig:app:reflect:gpt-3-5}/\ref{fig:app:reflect:gpt-4}.